\documentclass{JHEP3}
\usepackage{epsfig}

\newcommand{\be}{\begin{equation}}
\newcommand{\ee}{\end{equation}}
\newcommand{\la}{\langle}
\newcommand{\ra}{\rangle}
\newcommand{\proof}{\noindent {\bf Proof:} }
\newcommand{\statement}{\noindent {\bf Statement} }

\title{Universality of large $N$ phase transitions in 
Wilson loop operators in two and three dimensions}
\author{R. Narayanan
\\Department of Physics, Florida International University, Miami,
FL 33199, USA\\E-mail: \email{rajamani.narayanan@fiu.edu}}
\author{ H. Neuberger
\\ Rutgers University, Department of Physics and Astronomy,
Piscataway, NJ 08855, USA\\E-mail: \email
{neuberg@physics.rutgers.edu} }

\abstract {
The eigenvalue distribution of a Wilson
loop operator of fixed shape
undergoes a transition under scaling at infinite $N$.
We derive a large $N$ scaling function in
a double scaling limit of
the average characteristic polynomial associated with
the Wilson loop operator in two dimensional QCD.
We hypothesize that the transition in
three and four
dimensional large $N$ QCD are also in the same universality
class and provide a 
numerical test for our hypothesis in three dimensions.
}


\preprint{}

\begin{document}

\section{Introduction.}

Intuitively, when the closed curve defining a Wilson loop operator is 
uniformly scaled we expect a qualitative change to occur: for small loops
the parallel transport matrix round the loop is close to unity while for large 
loops this matrix should be as far from unity as possible, as a result of 
confinement. This is true in 2,3 and 4 Euclidean dimensions in pure YM, with 
gauge group $SU(N)$. 

As the scale of the loop is varied, the operator goes from being sensitive to 
short distance physics to being sensitive to long distance physics. Somewhere 
on the way it undergoes a crossover. In a previous 
paper~\cite{ourjhep} we put forward the hypothesis
that as $N$ increases the crossover narrows and becomes a 
phase transition at infinite $N$, in the sense usually applied to individual 
large matrices. The eigenvalue distribution for small Wilson loops is centered
around +1, and has a gap around -1. The gap is eliminated for large loops and
the eigenvalue distribution
covers the entire circle, becoming uniform for asymptotically large 
loops. Confinement means that the uniform
limit is approached with a correction that goes to zero 
exponentially in the square of the scale factor.  

The hypothesis is more than just asserting a transition in the sense that the
eigenvalue density has a point of non-analytic dependence on the scale parameter. 
The hypothesis also states that this phenomenon happens in 2, 3 and 4 Euclidean 
dimensions and that in all these dimensions the transitions are in the same 
universality class. For large $N$, close to the critical scale,
all the complicated dependence on loop shape comes in only through a finite
number of parameters, which are coefficients of terms dependent on 
sub-leading terms in $N$, of the form $N^{-\nu}$ with $\nu$ being 
universal exponents; further corrections in $\frac{1}{N}$ are less significant. 
The main exponent is related to the average eigenvalue spacing at -1 of the 
Wilson matrix close to criticality. The average 
spacing is then in between ${\cal O}(1)$, for a gap, 
and ${\cal O} ( N^{-1} )$, for nonzero eigenvalue density.  

The purpose of this paper is to test our hypothesis in continuum YM 
in 3 Euclidean dimensions by numerical Monte Carlo simulation on the lattice. 
Critical behavior induced by taking an extensive parameter to infinity is often 
tested by numerically confirming the presumed universal approach to 
the respective thermodynamic limit. In the case of ordinary second order phase 
transitions one may seek to identify a finite-size scaling function.  
Something similar needs to be done in the case of large $N$ transitions. 
The hypothesis we need to test says that the complicated 3 and 4 dimensional
cases have the same universal behavior as the exactly solvable 2 dimensional 
case. In two dimensions we know much about the approach to infinite $N$, 
where the transition has 
been established long ago. We refer to this transition as 
the DO transition,  after Durhuus and Olesen who discovered it~\cite{duol}.  
Since we are using lattice methods
to learn about continuum YM theory, we need to take the zero lattice spacing and
infinite volume limits. We work under the assumption that these limits interact
simply with the large $N$ limit. This is a standard assumption, and our results 
are consistent with it.  

We first derive the universal behavior of a specific observable related to the
Wilson matrix in two dimensions, where we work directly at infinite volume and
in the continuum.  
We then take
finite $N$ ``data'' arrived at by employing exact analytical formulas
and numerically check if this data
exhibits the universal asymptotic behavior in the crossover.  
With these tools
in hand we proceed to a numerical project in 
three dimensions, where in addition we need to handle 
statistical errors coming from the stochastic approximations used for the path
integrals and systematic errors having to do with not working directly in the
continuum and not at infinite volume. We exploit the relative 
ease to do simulations in three dimensions to make the statistical 
errors much smaller than absolutely necessary. Also, working at large $N$ 
reduces finite volume effects, leaving the approach to continuum as the main new
ingredient we need to get under control.  

\section{Two dimensions: basics.}

The Wilson loop matrix in YM on the infinite plane is given by
the product of many unitary matrices close to unity. Using methods first
introduced by Migdal~\cite{migdal}, the matrix associated with a 
curve that does not intersect itself is seen to be given by a product of a large
number of independently and identically distributed (i.i.d.) unitary matrices. These
unitary matrices are distributed in a small width around the unit matrix
and the probability distribution of the Wilson loop matrix depends on a
single parameter made out of the number of matrices and the width of their
distribution. This parameter is in one to one correspondence to the 
area enclosed by the loop in units of the gauge coupling constant.  

The multiplicative matrix model has been introduced by Janik and 
Wieczorek~\cite{janik} who employed a solution method
similar to that of Gopakumar and Gross~\cite{gogr}; 
we shall refer to it as the JW model.  
It's precise definition is: Let 
$U_i$ with $i=1,..,n$ be i.i.d. $N\times N$ unitary random matrices. 
$U_j=e^{i\epsilon H_j}$, where the hermitian matrix $H_j$ is either 
unconstrained or traceless
and distributed with a normalized probability density given by:
\be
P(U_j) = {\cal N} e^{-\frac{N}{2} {\rm Tr} H_j^2}
\ee
The parameter $\epsilon$ obeys $0<\epsilon << 1$ and the integer $n$ is large,
so that the product $\epsilon^2 n$ is finite. We shall take the limit
$n\to\infty$, $\epsilon\to 0$ with $t=n\epsilon^2$ kept fixed. $t$ is
related to the unit-less area mentioned above. The relation will be
made precise later on.  
The Wilson loop matrix is given by:
\be
W=\prod_{i=1}^n U_i
\ee

It turns out that the simplest gauge invariant observable made out
of $W$ which exhibits universal approach to critical behavior is the
average characteristic polynomial of $W$,
$\la\det(z-W)\ra$. 
The average characteristic polynomial is in one to one correspondence with
the set of traces of $W$ in all totally antisymmetric representations of
$SU(N)$ or $U(N)$. Nontrivial representations with zero $N$-ality do not enter.  

We shall derive integral and
polynomial expressions for
\be
Q_N(z,t)=\lim_{n\to\infty,\epsilon\to 0}
\left< \det(z-W)\right>|_{t=\epsilon^2n\ \ {\rm fixed}} 
\ee
that are valid for all $N$, separately for $SU(N)$ and for $U(N)$.
These results are used to find the $N\to\infty$ limit, find a critical loop 
size in that limit, and then zoom into the the vicinity of this infinite $N$ 
critical point.  This vicinity is described by a  ``double scaling limit'' of 
the average characteristic polynomial. The double scaling limit turns 
out to be identical for $SU(N)$ and for $U(N)$. 

\subsection{The average characteristic polynomial of the Wilson 
matrix.}
We will derive the integral relation
\be
Q_N(z,t)
 =\cases{
\sqrt{\frac{N\tau}{2\pi}}
\int_{-\infty}^\infty d\nu
e^{-\frac{N}{2}\tau\nu^2} 
\left[z-e^{-\tau\nu
-\frac{\tau}{2}
}
\right]^N & for $SU(N)$ \cr
\sqrt{\frac{Nt}{2\pi}}
\int_{-\infty}^\infty d\nu
e^{-\frac{N}{2}t\nu^2} 
\left[z-e^{-t\nu
-\frac{\tau}{2}
}
\right]^N & for $U(N)$ \cr}\label{inteqn}
\ee
where $\tau=t\left(1+\frac{1}{N}\right)$.
Given this relation, we can perform a binomial expansion
and then compute the integral to obtain the polynomial
relation
\be
Q_N(z,t)
 =\cases{
\sum_{k=0}^N 
\pmatrix{N\cr k\cr} z^{N-k} (-1)^k e^{-\frac{\tau k(N-k)}{2N}} &
for $SU(N)$ \cr
\sum_{k=0}^N 
\pmatrix{N\cr k\cr} z^{N-k} (-1)^k e^{-\frac{t k(N+1-k)}{2N}} &
for $U(N)$ \cr
}\label{poleqn}
\ee

Before we proceed to give the details of the derivation
of (\ref{inteqn}), we make some observations with regard
to the polynomial expressions for $Q_N(z,t)$.

\subsubsection{Heat-kernel measure for $W$ in the $SU(N)$ case.}

The definition of the $SU(N)$ random matrix ensemble
produces an evolution in ``time'' of the probability distribution of the
product matrix over the manifold of $SU(N)$. Invariance properties
and locality imply that, up to some rescaling of the variable $t$ to
a variable $\tau$, the
probability distribution of the product matrix,
$W$, will be given by the
heat-kernel for $SU(N)$:
\be
P(W,\tau ) dW = \sum_R d_R \chi_R (W) e^{-\tau C_2 (R)} dW
\ee
Here, $dW$ is the Haar measure on $SU(N)$, $R$ labels the irreducible
representations of $SU(N)$, $C_2(R)$ is the second order Casimir
in the representation $R$ and $\chi_R (W)$ is the character
of the representation $R$ evaluated on the matrix $W$, with the
convention that $\chi_R ({\mathbf 1})=d_R$ with $d_R$ being the
dimension of $R$. The normalization convention for the Casimir
operator are related to the scale freedom in $t$. 
The normalization of the Haar
measure $dW$ is such that the characters $\chi_R (W)$ are orthonormal with
respect to $dW$. Finally, the probability distribution is properly
normalized such that $\int P(W,\tau)dW =1$.

Let us now focus on the k-fold antisymmetric representations,
$k=1,..,N$ and label them by $k$. $C_2 (k) = A_N \frac {k(N-k)}{2N}$
and $d_k={N\choose k}$. We absorb $A_N$ in the definition of $\tau$.  
If the eigenvalues of $W$ are
$e^{i\theta_1},e^{i\theta_2},e^{i\theta_3},....,e^{i\theta_N}$, 
and we define the moments, $M_k(t)$, by
\be
M_k(t) = 
\langle
\sum_{1\le j_1 < j_2 < j_3....< j_k\le 
N}e^{i(\theta_{j_1}+\theta_{j_2}+...+\theta_{j_k})}\rangle,
\label{moments}
\ee
it follows that
\be
M_k(t)=\langle \chi_k (W) \rangle = 
d_k e^{-\tau C_2(k)} = {N\choose k} e^{-\frac{\tau k(N-k)}{2N}}
\label{cheqn}
\ee 

Next we note that 
\be
Q_N(z,t) = \langle \prod_{j=1}^N (z-e^{i\theta_j})\rangle
= \sum_{k=0}^N z^{N-k}
(-1)^k M_k(t) \label{momeqn}
\ee
and we are consistent with the SU(N) case in (\ref{poleqn})
if we use (\ref{cheqn}) above.
This consistency with a heat-kernel probability distribution
for $W$ provides a check of the derivation of
(\ref{inteqn}) in the $SU(N)$
case.

\subsubsection{$Q_N(z,t)$ does not self-average at finite
  $N$: $U(N)$ case.}

The moments $M_k$ defined in (\ref{moments}) are given by
\be
M_k(t)=\cases { e^{-\frac{\tau k(N-k)}{2N} } & for $SU(N)$ \cr
e^{-\frac{t k(N+1-k)}{2N} } & for $U(N)$ \cr}
\label{mkeqn}
\ee
as seen by matching (\ref{momeqn}) with (\ref{poleqn}).
We note that $M_k(t)=M_{N-k}(t)$ only for $SU(N)$
since $W$ and $W^\dagger$ are equally probable and $\det W=1$.

There is insufficient information contained in the moments $M$ to
determine the joint probability distribution of the $\theta_i$, or
even of the average resolvent,$\langle
\sum_{j=1}^N\frac{1}{z-e^{i\theta_j}}\rangle$.
Nevertheless, 
$Q_N(z,t)$ is a polynomial in $z$ and its zeros are determined
by the coefficients $M_k$. 
Obviously, for any fixed $W$, the $e^{i\theta_j}$ are
the zeros of $\det(z-W)$. Therefore, we expect the zeros of 
$Q_N(z,t)$ to represent in some manner the statistical
properties of the $e^{i\theta_j}$. For any finite $N$ there is no way
to obtain the exact marginal distribution of even just a single
eigenvalue of $W$ (``one point function'') from the average
characteristic polynomial.  However, in the large $N$ limit, this
often becomes possible.

It is obvious that in our case, at finite $N$, 
\be
\log \langle \det(z-W)\rangle \ne \langle\log \det(z-W)\rangle
\ee
for $U(N)$.
This is seen already by comparing the $z^0$ term on the two sides. 
That $\langle \det W\rangle =M_N(t)=e^{-\frac{t}{2}}$ 
follows from (\ref{momeqn}) and (\ref{mkeqn}).
It is easy to understand
the above result.
The probability of any one of
the hermitian matrices $H_j$ factorizes into a factor depending only
on the traceless part of $H_j$ ($SU(N)$ part)
and another depending just on ${\rm Tr} H_j$ ($U(1)$ part):
${\rm Tr} H^2 = {\rm Tr} (H-\frac{1}{N}{\rm Tr} H)^2 + \frac{1}{N} ({\rm Tr} H)^2$. 
$\det W$ only depends on the $U(1)$ part and since this is the
commuting part, we get
\be
\langle \det(W)\rangle=e^{-\frac{n\epsilon^2}{2}}=e^{-\frac{t}{2}}
\ee
On the other hand, $\langle \log \det(W)\rangle =0$.  

Obviously, this example is specific to $U(N)$ and would not hold in
the $SU(N)$ case.  The observation is nevertheless useful as it
provides an easy check of our derivation to follow.
This might be viewed as a $\frac{1}{N}$
effect, since one would expect $<\det (W)>\sim e^{-N(...)}$ at large
$N$. This is consistent with the difference between $U(N)$ and $SU(N)$
being of lower order in $\frac{1}{N}$.  However, since we are going to
look at a more subtle large $N$ limit, where we amplify a critical
regime introducing extra dependences on $N$ into some of the
parameters $z$ and $t$ (a ``double scaling'' limit) we need to be
careful about the distinction between $U(N)$ and $SU(N)$. Eventually
we shall see that the difference between $U(N)$ and $SU(N)$ indeed
does not matter as the exponents $\nu$ will be smaller than one. Thus,
the universal corrections to the singular behavior at the critical
point are larger than the $\frac{1}{N}$ correction differentiating
$U(N)$ from $SU(N)$.

\subsubsection{Zeros of $Q_N(z,t)$ and the
 Lee-Yang theorem~\cite{leeyang}.}

We have commented already that the information about the true
distribution of eigenvalues of the stochastic Wilson matrix is
represented by the average characteristic polynomial only in a
statistical sense, in that it would reproduce the moments contained in
the coefficients of the characteristic polynomial, but not necessarily
other spectral properties.  We show now that the roots of the average
characteristic polynomial in the case of $SU(N)$ are on the unit
circle, similarly to the roots of every instance of the random Wilson
matrix. This goes a long way toward justifying that the spectrum of
the average characteristic polynomial itself can be seen as an
approximation of the average spectrum of the Wilson loop matrix.

The polynomial expression for the $SU(N)$ in (\ref{poleqn}) 
is
\be
Q_N(z,t) = (-1)^N e^{-\frac{N\tau}{8}} (-z)^{\frac{N}{2}}
\sum_{k=0}^N {N\choose k} (-z)^{\frac{N}{2}-k} e^{\frac{\tau}{2N} 
(k-\frac{N}{2})^2}
\label{q-nn}
\ee 
Introduce now $N$ Ising spins,
$\sigma_i=\pm\frac{1}{2}, ~~i=1,...N$ and the magnetization
$M(\sigma)=\sum_{i=1}^N \sigma_i$. Then,
\be
M(\sigma)=\frac{N}{2}-k
\ee
where $k$ is the number of spins equal to $-\frac{1}{2}$ and
varies between $0$ and $N$. 

Taking into account the number of configurations with $k$ spins
equal to $-\frac{1}{2}$ we get:
\be
Q_N(z,t) = (-1)^N e^{-\frac{N\tau}{8}} (-z)^{\frac{N}{2}}
\sum_{\sigma_1,\sigma_2,...\sigma_N=\pm\frac{1}{2}} (-z)^{M(\sigma)} 
e^{\frac{\tau}{2N} M^2(\sigma)}
\ee 
The self interaction terms from the
magnetization squared can be extracted as a further prefactor. What
remains is the partition function of an Ising model on an $N$ vertex
graph where every vertex is connected to every other vertex.
\be
Z_N(z,t)=
Q_N(z,t)(-1)^N e^{\frac{(N-1)\tau}{8}} (-z)^{-\frac{N}{2}}=
\sum_{\sigma_1,\sigma_2,...\sigma_N =\pm\frac{1}{2}} 
e^{\ln(-z)\sum_i \sigma_i}
e^{\frac{\tau}{N} \sum_{i > j} \sigma_i \sigma_j}
\ee

The interaction is ferromagnetic for positive $\tau$ and there is a
complex external magnetic field $\log(-z)$. The conditions of the
Lee-Yang theorem~\cite{leeyang} are therefore fulfilled and all roots
of this partition function (and hence of the polynomial $Q_N(z,t)$) 
lie on
the unit circle.

This is a result about the finite $N$ average characteristic
polynomial, which holds for all $N$ in the $SU(N)$ case, but, as
expected and explained earlier, cannot and does not hold in the $U(N)$
case, where the circle on which the eigenvalues lie shrinks
exponentially with $t$.  

\subsubsection {Derivation of the integral representation for
$Q_N(z,t)$.}

We proceed to derive (\ref{inteqn}) through a series
of statements.
We will need the following external field
integrals over $H$ as part of our derivation. For $U(N)$ we have 
\be
\langle<e^{i\epsilon {\rm tr}(HX)}\rangle>=
e^{-\frac{\epsilon^2}{2N} {\rm tr}(X^2 )}
\label{unext}
\ee
and, for $SU(N)$ we have
\be
\langle<e^{i\epsilon {\rm tr}(HX)}\rangle>=
e^{-\frac{\epsilon^2}{2N} {\rm tr}(X^2) 
+\frac{\epsilon^2}{2N^2} (tr(X))^2}
\label{sunext}
\ee
The $SU(N)$ formula gives 
$\langle<e^{i\epsilon {\rm tr}(HX)}\rangle>=1$ 
for $X$ proportional to the unit matrix, as expected,
since $tr(HX)\propto {\rm Tr} (H)=0$. 

An essential tool in our derivation is a path integral
representation of the characteristic polynomial
which is set up in the following statement.

\statement I:

\be
\det (z-W) = 
\int \prod_{i=1}^n [ d\psi_id\bar\psi_i]
e^{\sum_{i=1}^n \left[ w\bar\psi_i \psi_i -\bar\psi_i U_i \psi_{i+1}\right]}
\ee
where $\bar\psi_i$, $\psi_i$ are Grassmann variables,
$z=w^n$, $W=\prod_{i=1}^n U_i$ and $\psi_{n+1}=\psi_1$.
As $n\to\infty$, $w\to 1$ while the complex
variable $z=w^n$ is held fixed. 

\proof 
This statement reflects the obvious gauge invariance of the Grassmann
system, in addition to a $Z(n)$ invariance under $w\to w e^{\frac{2\pi
i}{n}}$.  The proof is by recursion. One step in the recursion process
is
\be
\int d\psi_j d\bar\psi_j e^{w\bar\psi_j\psi_j -\bar\psi_jU_j\psi_{j+1}
-\bar\psi_n R_j \psi_j}
= w^N e^{-\bar\psi_n R_{j+1} \psi_{j+1}}
\ee
with 
\be
R_{j+1} = \frac{1}{w} R_jU_j.
\ee
Noting that $R_1=U_n$, we can repeat the single step above
to integrate out all Grassmann
variables except $\bar\psi_n$ and $\psi_n$ and obtain
the desired result of Statement I:
\begin{eqnarray}
\int \prod_{i=1}^n [ d\psi_id\bar\psi_i]
e^{\sum_{i=1}^n \left[ w\bar\psi_i \psi_i -\bar\psi_i U_i \psi_{i+1}\right]}
&=& w^{N(n-1)} \int d\psi_n d\bar\psi_n e^{w\bar\psi_n\psi_n
- \bar\psi_n R_n \psi_n}\cr
&=&w^{N(n-1)} \det (w - R_n)\cr
&=& \det (w^n - w^{n-1}R_n)\cr
&=& \det (z - W)
\end{eqnarray}

The derivation of (2.4) proceeds by first performing the average over $U_j$
followed by the integration over the Grassmann variables.
It will be useful to have an additional identity
in the form of another Grassmann integral
as stated below.

\statement II:

For $k>1$,
\begin{eqnarray}
e^{-\bar\psi Y^k \chi}
&=&\int \prod_{l=1}^{k-1} d\bar\eta^k_l d\eta^k_l
e^{-\sum_{l=1}^{k-1} \bar\eta^k_l \eta^k_l -\bar\psi Y \eta^k_1
+\sum_{l=1}^{k-2} \bar\eta^k_l Y \eta^k_{l+1} +\bar\eta^k_{k-1} Y\chi}\cr
&\equiv& \left < e^{-\bar\psi Y \eta^k_1
+\sum_{l=1}^{k-2} \bar\eta^k_l Y \eta^k_{l+1} +\bar\eta^k_{k-1} Y\chi}
\right>_\eta
\end{eqnarray}
 \proof For each $k>1$, we have $(k-1)$ pairs of Grassmann variables
denoted by $\bar\eta_l^k$ and $\eta_l^k$, $l=1,\cdots, (k-1)$.
in the above statement. The proof of this statement also works
by recursion. One step in the recursion process is
\be
\int d\bar\eta_l^k d\eta_l^k 
e^{-\bar\eta_l^k\eta_l^k -\bar\psi Y^l \eta_l^k
+\bar\eta_l^k Y \eta_{l+1}^k}
=  e^{-\bar\psi Y^{l+1} \eta_{l+1}^k}
\ee
Identifying $\eta_l^l=\chi$, we perform the above recursion
$(k-1)$ times, starting from $l=1$ to $l=(k-1)$ to obtain
the result of Statement II.

We can use the result of the Statement II to perform the
integral over $U$. We focus on one such integral in
the following statement.

\statement III:

\be
\left < e^{-\bar\psi U \chi} \right> = 
e^{-\bar\psi\chi}
\left <
\cases{
e^{-\frac{\epsilon^2}{2N}{\rm Tr}X^2+
\frac{\epsilon^2}{2N^2}({\rm Tr} X)^2} & for $SU(N)$ \cr
e^{-\frac{\epsilon^2}{2N}{\rm Tr}X^2} & for $U(N)$ \cr}
\right >_\eta
\ee
where
the matrix $X$ is
\be
X_{ij} =
\chi_i \bar\psi_j 
+\sum_{k=2}^{\infty} \frac{1}{(k!)^{1/k}}(\eta_1^k)_i\bar\psi_j
-\sum_{k=3}^\infty\sum_{l=1}^{k-2} \frac{1}{(k!)^{1/k}}
(\eta_{l+1}^k)_i(\bar\eta_l^k)_j
-\sum_{k=2}^\infty \frac{1}{(k!)^{1/k}}\chi_i(\bar\eta_{k-1}^k)_j
\ee
\proof

\begin{eqnarray}
\left < e^{-\bar\psi U \chi} \right> &=& 
\left < e^{-\bar\psi e^{i\epsilon H}\chi} \right > 
= \left < 
\prod_{k=0}^{\infty}
e^{-\bar\psi \left[\frac{i\epsilon H}{(k!)^{1/k}}\right]^k \chi}
\right > \cr
&=&
e^{-\bar\psi\chi}
\int \prod_{k=2}^\infty \prod_{l=1}^{k-1} d\bar\eta^k_l d\eta^k_l
e^{-\sum_{k=2}^\infty \sum_{l=0}^{k-1} \bar\eta_l^k \eta_l^k}
\left< e^{-i\epsilon{\rm Tr}HX}\right>\cr
&=&
e^{-\bar\psi\chi}
\left <
\cases{
e^{-\frac{\epsilon^2}{2N}{\rm Tr}X^2+
\frac{\epsilon^2}{2N^2}({\rm Tr} X)^2} & for $SU(N)$ \cr
e^{-\frac{\epsilon^2}{2N}{\rm Tr}X^2} & for $U(N)$ \cr}
\right >_\eta \label{stat3pr}
\end{eqnarray}
We have used Statement II to obtain the third equality in
(\ref{stat3pr}) and we have used (\ref{unext}) and
(\ref{sunext}) to obtain the
fourth equality in (\ref{stat3pr}).

We can now perform the integrals over the full set of
$\eta$ and $\bar\eta$ variables to get the result of
the following statement.

\statement IV

\be
\left < e^{-\bar\psi U \chi} \right> =
\sqrt{\frac{N}{2\pi}} \int_{-\infty}^\infty
d\lambda e^{-\frac{N}{2}\lambda^2}
e^{-\left[1-\lambda\epsilon\sqrt{1+\frac{u}{N}} 
- \frac{1}{2}\epsilon^2\left(1-\frac{u}{N^2}\right)\right]
\bar\psi\chi} 
\ee
with $u=0$ for $U(N)$ and $u=1$ for $SU(N)$.

\proof
In the limit of $n\to\infty$ and $\epsilon\to 0$, we can write
\be
\left <
e^{-\frac{\epsilon^2}{2N}{\rm Tr}X^2+
\frac{\epsilon^2}{2N^2}({\rm Tr} X)^2}\right>_\eta
= 
e^{-\frac{\epsilon^2}{2N}\left<{\rm Tr}X^2\right>_\eta+
\frac{\epsilon^2}{2N^2}\left< ({\rm Tr} X)^2\right>_\eta}
\label{dissun}
\ee
for $SU(N)$
and
\be
\left <
e^{-\frac{\epsilon^2}{2N}{\rm Tr}X^2}\right>_\eta
= 
e^{-\frac{\epsilon^2}{2N}\left<{\rm Tr}X^2\right>_\eta}
\label{disun}
\ee
for $U(N)$.
The connected correlators of the exponent ignored above
will result in terms of the form
\be
F(\zeta) = \epsilon^2f_1(\epsilon)\zeta+\epsilon \sum_{k=2}^\infty
f_k(\epsilon)\zeta^k 
\label{extraf}
\ee
where $\zeta=\epsilon\bar\psi\chi$ and $f_k(\epsilon)$,
$k=1,\cdots\infty$ have a power series expansion in $\epsilon$
with only non-negative powers.
That the terms can only depend on the fermion bilinear $\bar\psi\chi$
is evident from symmetry arguments. 
Since $X$ appears with one power of $\epsilon$
on the left hand side of (\ref{dissun}) and (\ref{disun}), we can
associate a $\sqrt{\epsilon}$ with each fermion. Every term
in the connected correlator that contributes
should have at least one term
of the form $\bar\eta_l^k\eta_l^k$ that got integrated.
This gives at least one
extra power of $\epsilon$. If the connected correlator
has to result in $\zeta$, the term should have at least
two terms of the form $\bar\eta_l^k\eta_l^k$ since
the relevant terms comes from $\left({\rm Tr} X^2\right)^2$, 
$\left({\rm Tr} X\right)^4$, ${\rm Tr} X^2\left({\rm Tr} X\right)^2$
or higher powers of $X$. 
The extra powers of $\epsilon$ in (\ref{extraf})
result in the vanishing of
this term in the $\epsilon\to 0$ and $n\to\infty$
limit.

Even though $X$ has an infinite number of terms, there are
only two terms in $\left < {\rm Tr} X^2\right >_\eta$ 
and  two terms in $\left < ({\rm Tr} X)^2\right >_\eta$:
\begin{eqnarray}
\left < {\rm Tr} X^2\right >_\eta &=& -(\bar\psi\chi)^2 -N \bar\psi\chi \cr
\left < ({\rm Tr} X)^2\right >_\eta &=& (\bar\psi\chi)^2 - \bar\psi\chi
\label{trx2}
\end{eqnarray}
Inserting (\ref{trx2}) into (\ref{dissun}) and
(\ref{disun}) and the result into Statement III
gives us
\be
\left < e^{-\bar\psi U \chi} \right> 
= 
e^{-\bar\psi\chi}
e^{\frac{\epsilon^2}{2N}\left(1+\frac{u}{N}\right)
(\bar\psi\chi)^2+\frac{\epsilon^2}{2}\left(1-\frac{u}{N^2}\right)
\bar\psi\chi};
\label{oe2}
\ee
with $u=0$ for $U(N)$ and $u=1$ for $SU(N)$.

Since,
\be
\sqrt{\frac{N}{2\pi}}
\int_{-\infty}^\infty d\lambda e^{-\frac{N}{2}\lambda^2 
+\lambda\epsilon\sqrt{1+\frac{u}{N}}\bar\psi\chi} 
= e^{\frac{\epsilon^2}{2N}\left(1+\frac{u}{N}\right)(\bar\psi\chi)^2}
\ee
the result in (\ref{oe2}) reduces to statement IV.

Had we kept the term 
$F(\zeta)$ in (\ref{extraf}), we 
can change the factor $e^{-{\frac{N}{2}\lambda^2}}$ in the integrand
depending on the auxiliary fields $\lambda$ to
$e^{-{\frac{N}{2}\lambda^2}}(1+P(\epsilon,\lambda))$ so as to
reproduce those terms.
Alternatively, one
can introduce an auxiliary field capturing the entire 
$F(\zeta)$
dependence using an inverse Laplace transform and interpreting it
perturbatively (that is not worrying about the convergence of the
$\lambda$ integration, as anything beyond quadratic order is assumed
to get expanded and truncated according to the power of
$\epsilon$). In either case one gets extra terms that will vanish in
the correlated, large $n$ -- small $\epsilon$, limit.  

Now, we can use Statement IV to perform each $U_i$ integral appearing
in the expression for $\det(z-W)$ in Statement I resulting in
the following statement.

\statement V

\begin{eqnarray}
\left< \det(z-W)\right> =&&\left(\frac{N}{2\pi}\right)^\frac{n}{2}
\int \prod_{i=1}^n d\lambda_i e^{-\frac{N}{2}\sum_{i=1}^n
\lambda_i^2} \cr
&&\int \prod_{i=1}^n [ d\psi_id\bar\psi_i]
e^{\sum_{i=1}^n \left[ w\bar\psi_i \psi_i -
\left[1-\lambda_i\epsilon\sqrt{1+\frac{u}{N}}
-\frac{\epsilon^2}{2}\left(1-\frac{u}{N^2}\right)\right]\bar\psi_i\psi_{i+1}\right]}
\end{eqnarray}

We can now perform the Grassmann integrals exactly following
the proof of Statement I for $1\times 1$ matrices. The result
is stated below.

\statement VI

\begin{eqnarray}
\left< \det(z-W)\right> &=&\left(\frac{N}{2\pi}\right)^\frac{n}{2}
\int \prod_{i=1}^n d\lambda_i e^{-\frac{N}{2}\sum_{i=1}^n
\lambda_i^2} \cr
&&\left[z-\prod_{i=1}^n \left[1-\lambda_i\epsilon\sqrt{1+\frac{u}{N}}
-\frac{\epsilon^2}{2}\left(1-\frac{u}{N^2}\right)\right]
\right]^N
\end{eqnarray}

We are now set to proof the main result, namely, (\ref{inteqn}).
We start by exponentiating the term inside the product
of statement VI.
One needs to
take into account then a term of order $\epsilon^2\lambda_i^2$ which
makes a finite contribution. This term is inserted into the
exponentiated form in such a manner that the agreement between the
exponentiated expression and the original one holds also at order $\epsilon^2$.  
\begin{eqnarray}
\left< \det(z-W)\right> &=&\left(\frac{N}{2\pi}\right)^\frac{n}{2}
\int \prod_{i=1}^n d\lambda_i e^{-\frac{N}{2}\sum_{i=1}^n
\lambda_i^2} \cr
&&\left[z-e^{-\epsilon\sqrt{1+\frac{u}{N}}\sum_i\lambda_i
-\frac{n\epsilon^2}{2}\left(1-\frac{u}{N^2}\right)
-\frac{\epsilon^2}{2}\left(1+\frac{u}{N}\right)\sum_i \lambda_i^2
}
\right]^N
\end{eqnarray}
Let $\Lambda=(\lambda_1,\lambda_2,\cdots,\lambda_n)$.
let $r_1,r_2,\cdots,r_n$ be an orthonormal basis with
$r_1=\frac{1}{\sqrt{n}}(1,\cdots,1)$.
Finally, let $r_i\cdot\Lambda=\xi_i$. Then, using $t=n\epsilon^2$,
\be
\left< \det(z-W)\right> =\left(\frac{N}{2\pi}\right)^\frac{n}{2}
\int \prod_{i=1}^n d\xi_i e^{-\frac{N}{2}\sum_{i=1}^n
\xi_i^2} 
\left[z-e^{-\sqrt{t}\sqrt{1+\frac{u}{N}}\xi_1
-\frac{t}{2}\left(1-\frac{u}{N^2}\right)
-\frac{\epsilon^2}{2}\left(1+\frac{u}{N}\right)\sum_i \xi_i^2
}
\right]^N
\ee
Now, let $\xi_1=\sqrt{t}\mu$ and $\xi_k=\sqrt{n}\mu_k$ for $k=2,\cdots,n$.
Then, again using $t=n\epsilon^2$,
\begin{eqnarray}
\left< \det(z-W)\right> =&&\left(\frac{N}{2\pi}\right)^\frac{n}{2}
\sqrt{t} n^{\frac{n-1}{2}}\int d\mu 
\int \prod_{i=2}^n d\mu_i e^{-\frac{N}{2}t\mu^2-\frac{N}{2}n\sum_{i=2}^n
\mu_i^2} \cr
&& \left[z-e^{-t\sqrt{1+\frac{u}{N}}\mu
-\frac{t}{2}\left(1-\frac{u}{N^2}\right)
-\frac{t\epsilon^2}{2}\left(1+\frac{u}{N}\right)\mu^2
-\frac{t}{2}\left(1+\frac{u}{N}\right)\sum_{i=2}^n \mu_i^2
}
\right]^N
\end{eqnarray}
Next, we go to polar coordinates in $\mu_i$, $i=2,\cdots,n$ and
set $r^2=\sum_{i=2}^n \mu_k^2$. Then we get,
\begin{eqnarray}
\left< \det(z-W)\right> =&&\left(\frac{N}{2\pi}\right)^\frac{n}{2}
\sqrt{t} n^{\frac{n-1}{2}} \frac{2\pi^{\frac{n-1}{2}}}
{\Gamma\left(\frac{n-1}{2}\right)}
\int_{-\infty}^\infty d\mu 
\int_0^\infty dr r^{n-2}
e^{-\frac{N}{2}t\mu^2-\frac{N}{2}n r^2} \cr
&& \left[z-e^{-t\sqrt{1+\frac{u}{N}}\mu
-\frac{t}{2}\left(1-\frac{u}{N^2}\right)
-\frac{t\epsilon^2}{2}\left(1+\frac{u}{N}\right)\mu^2
-\frac{t}{2}\left(1+\frac{u}{N}\right)r^2
}
\right]^N
\end{eqnarray}
For large $n$, we can perform a saddle point calculation of
the $r$ integral.
To leading order in $n$ the saddle point is at 
$r_c=\sqrt{\frac{1}{N}}$ and we get
\begin{eqnarray}
\left< \det(z-W)\right> =&&\left(\frac{N}{2\pi}\right)^\frac{n}{2}
\sqrt{t} n^{\frac{n-1}{2}} \frac{2\pi^{\frac{n-1}{2}}}
{\Gamma\left(\frac{n-1}{2}\right)}
N^{-\frac{n}{2}}e^{-\frac{n}{2}}\sqrt{\frac{\pi}{nN}}N
\int_{-\infty}^\infty d\mu 
e^{-\frac{N}{2}t\mu^2} \cr
&& \left[z-e^{-t\sqrt{1+\frac{u}{N}}\mu
-\frac{t}{2}\left(1+\frac{1}{N}\right)
-\frac{t\epsilon^2}{2}\left(1+\frac{u}{N}\right)\mu^2
}
\right]^N
\end{eqnarray}
Now we take the limit, $n\to \infty$, $\epsilon\to 0$ for a fixed $t$
and we get
\be
\left< \det(z-W(t))\right> =\sqrt{\frac{Nt}{2\pi}}
\int_{-\infty}^\infty d\mu 
e^{-\frac{N}{2}t\mu^2} 
\left[z-e^{-t\sqrt{1+\frac{u}{N}}\mu
-\frac{t}{2}\left(1+\frac{1}{N}\right)
}
\right]^N
\ee
Finally, we define
$\tau=t\left(1+\frac{1}{N}\right)$ and
$\nu=\frac{\mu}{\sqrt{1+\frac{u}{N}}}$.
Then the above equation reduces to
(\ref{inteqn}).

We note that $U(N)$ and $SU(N)$ become 
indistinguishable in the large $N$ double scaling limit we shall later
employ. We will restrict ourselves to the $SU(N)$ case at finite $N$.

Since we are interested in the large $N$ limit, we will not
distinguish between $\tau$ and $t$ and we will set
\be
Q_N(z,t)
 =
\sqrt{\frac{Nt}{2\pi}}
\int_{-\infty}^\infty d\nu
e^{-\frac{N}{2}t\nu^2} 
\left[z-e^{-t\left(\nu+\frac{1}{2}\right)}\right]^N
\label{qnzt}
\ee
for all the discussion to follow.
If we need to compare to continuum two dimensional
YM, we should keep in mind that, on the basis of a comparison to a 
heat-kernel formula for the average characteristic polynomial, 
it is $\tau$ that is related directly to the inverse 't Hooft
coupling, not $t$. In other words, there is a factor of $1+\frac{1}{N}$
in the relationship between the parameter $t$ of the JW model and
the inverse 't Hooft coupling in the standard notation for the
dimensionless area in two dimensional $SU(N)$ YM theory.  

\subsubsection{The average characteristic 
polynomial for negative areas.}

The average characteristic polynomial depends on $t$ in an analytic
manner.  In particular, it is interesting to consider the case $t\le
0$. The conditions for applying the Lee-Yang theorem no longer hold,
as the interaction has become anti-ferromagnetic. Explicit examples
show that all roots of the average characteristic polynomial are real
and negative.  On the other hand, it remains true for any $t$ that if
$z$ is a root so are $\frac{1}{z}, z^*, \frac{1}{z^*}$. These
symmetries are consistent with restricting all roots to the unit
circle or to the positive or negative portions of the real axis. Thus,
the symmetries alone do not tell much.

The case of $t\le 0$ corresponds to imaginary $\epsilon$, since
$t=n\epsilon^2$.  Imaginary $\epsilon$ 
corresponds to a complex Wilson matrix
obtained by multiplying i.i.d. hermitian matrices
close to identity. One would expect this Wilson matrix to have a
spectrum covering a region of the complex plane in the stochastic
sense. In this case we see that the roots of the average
characteristic polynomial carry little information about the spectral
properties of the Wilson matrix. This makes it clear why we carried
out various checks to convince ourselves that the average
characteristic polynomial was a useful observable for $t\ge 0$,
when the matrices that get multiplied are unitary.

\subsection{The large $N$ phase transition in $Q_N(z,t)$.}

We end up concluding that the average characteristic polynomial,
$Q_N(z,t)$, at infinite $N$ should reproduce the DO phase
transition. In other words, one can replace the average of a logarithm
by the logarithm of the average; this is somewhat analogous to a
self averaging result proved by Berezin in 1972~\cite{berezin}.

We check that the DO transition is captured by the average
characteristic polynomial by comparing our result to that of
~\cite{janik}, who showed that the multiplicative matrix model has the
DO phase transition using different methods, not involving the average
characteristic polynomial, but rather accessing the resolvent
$\lim_{N\to\infty}\frac{1}{N} \la tr
\frac{1}{z-W}\ra$ directly.  
At infinite $N$, there is no distinction between $SU(N)$ and $U(N)$. 
We take the large $N$ limit by finding the saddle point in $\mu$ that
controls the integral; at the saddle point, $\nu=\lambda(t,z)$. 
\be
\frac{1}{N} \log Q_N(z,t)
=-\frac{1}{2N}\log\left[1+ t\left(\lambda^2+\lambda\right)\right] 
+\log \left ( z-
e^{-t (\lambda+\frac{1}{2})}\right ) -\frac{t}{2}\lambda^2
\ee
Here, $\lambda$ solves:
\be
\lambda=\lambda(t,z)=\frac{1}{ze^{t(\lambda+\frac{1}{2})}-1}
\ee

To get the resolvent of $W$ we take a derivative with respect to $z$. 
Only the explicit $z$ dependence on the right hand side matters, since
the expression is stationary with respect to variation in $\lambda$. 
We need to interchange the matrix averaging and the logarithm (a procedure 
we now have reason to believe will be valid in the limit of infinite $N$)
at fixed $t$ and $z$. The interchange can be viewed as a
version of large $N$ factorization, but now extended to a quantity
that has an exponential dependence on $N$. This ``self-averaging''
property may also hold in the double scaling limit we shall introduce
later, because violations of factorization would typically be (in
view of the new type of observable) of order $\frac{1}{N}$
while the double scaling limit will be seen to
add some dependencies in the couplings which are of slightly lower
order coming in via factors of $\frac{1}{N^\nu}$ with
$\nu=1/2,3/4$.  

The expression for the resolvent in the 
large $N$ limit is:
\be
G=\frac{1}{N} \langle {\rm Tr} \frac{1}{z-W}\rangle = 
\frac{1}{z-e^{-t\left(\lambda+\frac{1}{2}\right)}}\ee 

JW define a function $f(t,z)$ by:
\be
f(t,z)=zG(z,t)-1
\ee
and it is easy to see that $f$ and $\lambda$ are the same.
The equation for $\lambda$ can be rewritten as:
\be
z\lambda=(1+\lambda) e^{-t\left(\lambda+\frac{1}{2}\right)}
\ee
leading to equation (17) in~\cite{janik}.  This allowed us to bypass
the usage of the $S$-transform trick of~\cite{voicu} employed in
~\cite{janik}. We needed to bypass the usage of the $S$-transform
trick, because we need the universal smoothed out behavior at
asymptotically large, but not infinite, $N$, and the $S$-transform
procedure has no known extension away from the infinite $N$ limit.

We conclude that the average characteristic polynomial has a critical
point at infinite $N$ at $t=4$, which is the location of the DO phase
transition. The transition is reflected by the behavior around $z=-1$,
which is where the gap in the eigenvalue is first opened.

\section{The double scaling limit.}

We wish to zoom into the region close to $z=-1$ when $t$ is close to its 
critical value of $4$. Our previous discussion has led us to conclude that a 
good quantity to look at is the derivative of the logarithm of the average
characteristic polynomial with respect to $z$ at $z=-1$. It simplifies matters 
to focus on the real $z$ axis.

\subsection{ General structure: dimensions 2, 3, 4.}

We set $z=e^y$ and define a function $F(y)$ from
$Q_N(z,t)$ that is explicitly even in $y$:
\be
F(y)=e^{-\frac{Ny}{2}} \left (-1\right )^N Q_N( -e^y
  , t)=
=\langle \det\left
  (e^{\frac{y}{2}}+e^{-\frac{y}{2}}W \right )\rangle
\label{fofy}
\ee
We have suppressed the dependence on $t$ in the function $F(y)$.

We now introduce some new variables and notations:
\be
\Upsilon=\tanh\frac{y}{2},~~~A=\frac{1-W}{1+W} =-iM
\ee
$A$ is anti hermitian and $M$ is hermitian. If $e^{i\theta}$ is an
eigenvalue of $W$, $-i\tan\frac{\theta}{2}$ is the corresponding
  eigenvalue of $A$ and $\tan\frac{\theta}{2}$ of $M$. $A,M$ become
    singular when the gap in the eigenvalue spectrum of $W$
    closes. The inverse transformation to $W$ is:
\be
W=\frac{1-A}{1+A}=\frac{1+iM}{1-iM},~~~~1+W=\frac{2}{1+A}=\frac{2}{1-iM} 
\ee

The density of eigenvalues of $W$ is denoted by $\rho_N(\theta)$,
normalized by:
\be
\int_{-\pi}^\pi \rho_N(\theta ) \frac{d\theta}{2\pi}=N
\ee

$F(y)$ can be evaluated by Monte Carlo simulations in dimensions
higher than 2.  
\begin{eqnarray}
&F(y)=\left (2\cosh\frac{y}{2}\right )^N \langle \det\left (\frac
  {1}{1+A}\right ) \det(1+\tanh\frac{y}{2}\;\; A) \rangle=\cr & \left
  ( \frac{4}{1-\Upsilon^2}\right )^{\frac{N}{2}} \langle \det\left
  (\frac{1}{1+A}\right )  e^{-\sum_{n=1}^\infty (-1)^n \frac{\Upsilon^n}{n} {\rm Tr} 
A^n } \rangle\end{eqnarray}

This equation is still exact. For each $A$, $\det(1-A)=\det(1+A)$ on
account of the $SU(N)$ condition $\det W=1$. Since, in addition, 
the probability for an $A$ equals
that for a $-A$, $F(y)$ (which also depends on the loop and on the gauge
coupling) is even in $y$ (and, evidently then, in $\Upsilon$). This can be made 
explicit:\be
F(y)= \left
  ( \frac{4}{1-\Upsilon^2}\right )^{\frac{N}{2}} \langle \det\left
  (\frac{1}{1+A}\right )  e^{-\sum_{k=1}^\infty \frac{\Upsilon^{2k}}{2k}
  {\rm Tr} A^{2k} } \; \cosh\left ( \sum_{k=1}^\infty
  \frac{\Upsilon^{2k-1}}{2k-1} {\rm Tr} A^{2k-1} \right ) \rangle
\ee
From the above equation one can derive exact expressions for the
coefficients $F_k$ in $F(y)=\sum_{k=0}^\infty F_{k} \Upsilon^{2k}$.

If the joint
distribution of all the eigenvalues of $A$ were known one could
replace the averaging brackets on the right hand side by an integral
over all eigenvalues weighted by that distribution. If we apply large
$N$ factorization, the right hand side simplifies considerably, and
one is able to write it just in terms of the single eigenvalue distribution
of $A$. From previous discussions we feel it is fine to assume that 
large $N$ factorization holds in this case. 

If we apply large $N$ factorization, and use ${\rm Tr} A^{2k+1}=0$ for
integer $k$, we obtain:
\be
F_{\rm factorized} (\Upsilon) = \left ( \frac{4}{1-\Upsilon^2}\right
)^{\frac{N}{2}}
\frac{1}{\sqrt{\langle \det(1-A^2)\rangle}}\exp\left (
  -\sum_{k=1}^\infty \frac{\Upsilon^{2k}}{2k}\langle {\rm Tr} A^{2k} \rangle
\right )
\ee
In terms of $M$, we have:
\be
F_{\rm factorized} (\Upsilon) = \left ( \frac{4}{1-\Upsilon^2}\right
)^{\frac{N}{2}}
\frac{1}{\sqrt{\langle \det(1+M^2)\rangle}}\exp\left (
  \sum_{k=1}^\infty (-1)^{k-1} \frac{\Upsilon^{2k}}{2k}\langle {\rm Tr} M^{2k} \rangle
\right )
\ee

Let the eigenvalues of $M$ be denoted by $\lambda$. The 
eigenvalue density in $\theta$, $\rho_N (2\arctan y )$, which we now denote by 
an abuse of notation as $\rho_N (\lambda )$, is normalized by:
\be
\frac{1}{\pi}\int \frac{d\lambda}{1+\lambda^2} \rho_N(\lambda) = N
\ee
$\rho_N(\lambda )$ is an even function: $\rho_N(\lambda )=\rho_N(-\lambda )$.
When $\theta$ is close to $\pm\pi$, $\lambda$ goes to $\pm \infty$.  
The critical regime around $\theta\approx \pm\pi$ we are interested in
has been mapped to $\lambda\to\pm\infty$.  The eigenvalue spacing in
$\theta$ goes as the spacing in $\frac{1}{\lambda}$ in the large
$|\lambda |$ regime.

Let us assume a very large, but finite $N$. If $W$ is gap-less at
infinite $N$ at $-1$, $\rho_N(\lambda) \sim cN,~c>0$ as
$\lambda\to\pm\infty$. If $W$ has a gap in the infinite $N$ limit,
$\rho_N(\lambda)  \sim e^{-c^\prime N}, ~c^\prime >0$ as
$\lambda\to\pm\infty$. At the critical point when the gap just closes
at $\pm\pi$, $\rho_N(\lambda)\sim c^{\prime\prime} N
|\lambda|^{-\frac{1}{3}}$~\cite{janik}, as $\lambda\to\pm\infty$.  

If we now take the infinite $N$ limit, $\frac{1}{N}
\rho_N$ converges point-wise to a function $\rho_\infty$ that has
compact support if there is a gap, infinite support with regular
behavior at infinity if there is no gap and infinite support with a
singular behavior at infinity if we are exactly at criticality.  
One can then re-express the logarithmic derivative of the factorized $F$
\be
\frac{1}{N} \frac{\partial}{\partial \Upsilon} \log F_{\rm factorized}
(\Upsilon)=\frac{\Upsilon}{1-\Upsilon^2}+\frac{\Upsilon}{N}\langle 
{\rm tr}
\frac{M^2}{1+\Upsilon^2 M^2}\rangle
=\frac{\Upsilon}{N(1-\Upsilon^2)}
\langle{\rm tr} \frac{1+M^2}{1+\Upsilon M^2}\rangle
\ee
in terms of $\rho_\infty(\lambda)$ 
as
\be
\lim_{N\to\infty}
\frac{1}{N} \frac{\partial}{\partial \Upsilon} \log F_{\rm factorized}
(\Upsilon)
=\frac{1}{1-\Upsilon^2}\frac{\Upsilon}{|\Upsilon|}
\frac{1}{\pi}
\int_{-\infty}^\infty d\lambda \frac{\rho_\infty
\left(\frac{\lambda}{|\Upsilon|}\right)}{1+\lambda^2}
\label{ffact}
\ee

\subsubsection{Heuristic picture of the large $N$ 
phase transition.}\label{heur}
The determinant $\det(z-W)$ can be
thought of as the exponent of a sum of $N$ logarithms, one term for each
eigenvalue. It is then the exponent of the two dimensional electrostatic 
potential created by $N$ charges located at the zeros of the characteristic 
polynomial. These zeros are on the unit circle and we can look at the potential 
in the vicinity of the point $-1$ on this circle. There are two extreme cases: 
all charges are located at $+1$ or, the total charge is uniformly distributed on 
the circle. 

For the extreme case where all charges are located at $+1$,
$\rho_\infty(\lambda) = \pi \delta(\lambda)$. Inserting this
into (\ref{ffact}) results in
\be
\frac{1}{N} \frac{\partial}{\partial \Upsilon} \log F_{\rm factorized}
(\Upsilon)=\frac{\Upsilon}{1-\Upsilon^2}
\ee

For the other extreme case
of a uniform distribution of charges on the unit circle, 
$\rho_\infty (\lambda )= 1$. Inserting this
into (\ref{ffact}) results in
\be
\frac{1}{N} \frac{\partial}{\partial \Upsilon} \log F_{\rm factorized}
(\Upsilon)=\frac{\epsilon(\Upsilon)}{1-\Upsilon^2}
\ee

Recalling that
$\frac{\partial}{\partial\Upsilon}=\frac{2}
{1-\Upsilon^2}\frac{\partial}
{\partial y}$
we conclude that
\be
\frac{1}{N} \frac{\partial}{\partial y} \log F
(y)=\cases{
\frac{1}{2}\tanh\frac{y}{2} & for all charges at $+1$\cr
\frac{1}{2}\epsilon(y) & for uniform distribution of charges on the
unit circle \cr}
\ee
For a charge distribution that is critical, 
$\frac{1}{N} \frac{\partial}{\partial y} \log F
(y)$ goes as $y^{\frac{1}{3}}$ as $y$ goes to zero.
If we now
rescale the $y$ variable by $N^{\frac{3}{4}}$, defining $y
=\frac{\xi}{N^{\frac{3}{4}}}$,
$\frac{1}{N} \frac{\partial}{\partial y} \log F
(y)$ becomes
of order $N^{-\frac{1}{4}}$ for fixed $\xi$.

The double scaling limit will smooth out the non-analyticity at $y=0$ which
we exhibited explicitly above for the case of 
a uniform distribution.  
At infinite $N$, there will always be a non-analyticity at $y=0$ if
the eigenvalue distribution has no gap, whether the
distribution is uniform or not. The jump is proportional
to the density of eigenvalues of $W$ at $z=-1$, $\rho(\pi)$.  
On the other hand, when there is a gap, the behavior
at $y=0$ is smooth.

Up to a few non-universal parameters, the double-scaling limit
captures the universal content of the non-analyticity. 
Neither of the two extreme limits that we have seen above,
namely, a delta function and a uniform distribution, are
necessary to be attainable in a particular model, 
for the transition represented by the non-analyticity 
we have seen to take place and be universally described by the
scaling limit. 

To match the scaling limit to the data of a particular
physical realization, some parameters will need to be fit. 
For a large physical loop, one expects an almost uniform distribution.  
Suppose now that we have a distribution that is almost uniform, with a 
small deviation from uniformity proportional to $\cos\theta$. In
terms of $\lambda$,
\be
\rho(\lambda)=1+\delta \frac{1-\lambda^2}{1+\lambda^2}
\ee
Inserting this
into (\ref{ffact}) results in
\be
\frac{1}{N} \frac{\partial}{\partial \Upsilon} \log F_{\rm factorized}
(\Upsilon)=\frac{\epsilon(\Upsilon)}{1-\upsilon^2}\left[ 1 + \delta
\frac{|\Upsilon|-1}{|\Upsilon|+1}\right]
\ee
In terms of the variable $y$, the result is
\be
\frac{1}{N} \frac{\partial}{\partial y} \log F(y)=
\frac{\epsilon(y)}{2}\left [ 1 - \delta e^{-|y|}\right]
\ee

For a large loop, when the deviation of the eigenvalue distribution
from uniformity is small and determined by the string tension times
the area $t$, we have $\delta\propto e^{-\sigma t}$.  Positive and
negative $y$ values are related by a Z(2) symmetry.  The result is odd
in $y$ and undergoes a discontinuous change as $y$ goes through 0.
Taking a first derivative with respect to $y$ of the above equation,
we see that the area law term dominates for $y\ne 0$.

\subsection{ Structure in two dimensions.}

Inserting (\ref{qnzt}) into the definition of $F(y)$ in
(\ref{fofy}) yields
\be
F(y)=e^{-\frac{Ny}{2}} \left (-1\right )^N Q_N( -e^y
  , t)=2^N e^{-\frac{Nt}{8}}\sqrt{\frac{Nt}{2\pi}} Z_N (y,t),
\ee
where,
\be
Z_N(y,t) =\int dx \;\; e^{N[\log(\cosh \frac{y+tx}{2}) -\frac{1}{2} tx^2 ]}
\label{znyt}
\ee

We now extract from $Z_N$ the same factor we had extracted from $F$:
\be
Z_N(y,t) =\left(\cosh\frac{y}{2}\right )^N \int dx \;\; 
e^{ N\left [\log \left (\frac{\cosh \frac{y+tx}{2}}{\cosh\frac{y}{2}}\right ) 
-\frac{1}{2}tx^2\right ] }
\ee
Expanding the hyperbolic cosine of the sum in the exponent, we get:
\be
Z_N(y,t) =\left(\cosh\frac{y}{2}\right )^N \int dx \;\;e^{ N\left
    [\log \left ( 1+\tanh\frac{y}{2} \tanh\frac{tx}{2}\right ) 
-\frac{2}{t} \left ( \frac{tx}{2} \right )^2 
-\frac{1}{2}\log\left ( 1-\tanh^2 \frac{tx}{2}\right ) \right ] }
\ee

We change variables of integration from $x$ to
$v=\tanh\frac{xt}{2}$. The inverse transformation is 
$\frac{xt}{2}=-\frac{1}{2}\log\frac{1-v}{1+v}=\sum_{k=0}^\infty
\frac{v^{2k+1}}{2k+1}$.  

\be
Z_N(y,t)=\left(\cosh\frac{y}{2}\right )^N \frac{2}{t} \int_{-1}^1 
\frac{dv}{1-v^2}\;e^{ N\left
    [\log \left ( 1+\tanh\frac{y}{2} v \right ) -\frac{1}{2t} \left (
      \log\frac{1-v}{1+v} \right )^2 -\frac{1}{2}\log\left ( 1-v^2 \right ) 
\right ] }\ee

From now on, the integration over $v$ will be implicitly understood to run from
$-1$ to $+1$. We also introduce the parameter $\Upsilon=\tanh\frac{y}{2}$
with the understanding that $\Upsilon$ is real and that $|\Upsilon|\le 1$. Expanding the exponent in $v$ we have 
\begin{eqnarray}
Z_N(y,t) &=& \left ( \frac{1}{1-\Upsilon^2}\right
)^{\frac{N}{2}}\frac{2}{t} \int \frac{dv}{1-v^2}\times\cr
&&e^{ N\left [ \sum_{n=1}^\infty (-1)^{n-1}\frac{1}{n}\Upsilon^n v^n
-\frac{2}{t}v^2\left (\sum_{k=0}^\infty \frac{1}{2k+1} v^{2k}
\right)^2 +\frac{v^2}{2}\left (\sum_{k=0}^\infty
  \frac{1}{k+1}v^{2k}\right ) \right ] }\cr
&=&\left ( \frac{1}{1-\Upsilon^2}\right
)^{\frac{N}{2}}\frac{2}{t} \int \frac{dv}{1-v^2} \times\cr
&&e^{ N\left [ \left ( (\Upsilon v) -\frac{1}{2}(\Upsilon v)^2
    +\frac{1}{3}(\Upsilon v)^3+\dots \right ) +\left (
    (\frac{1}{2}-\frac{2}{t} ) v^2 +\frac{1}{4} v^4
    -\frac{4}{3}\frac{v^4}{t}\dots \right) \right ] }
\label{serieseq}
\end{eqnarray}

The critical point is at $t=4$, where the coefficient of the term
$Nv^2$ vanishes (the term of order $v^2$ that has a coefficient of
order 1 does not matter, as we are interested in the large $N$
critical point). The double scaling limit is defined so that the
highest power of $v$ (without a factor of $\Upsilon$) is 4. This means
that the integration variable $v$ will be conveniently redefined as
\be
v=\left (\frac{12}{N}\right )^{\frac{1}{4}} u
\ee
To keep a $\Upsilon$ dependence we need rescale $\Upsilon$ so that the
variable $\xi$ below is kept fixed as $N\to\infty$.  
\be
\Upsilon=\frac{\xi}{12^{\frac{1}{4}}\; N^{\frac{3}{4}}}
\ee
To keep the $v^2$ dependence we need to keep $t$ close to $4$, writing
\be 
\frac{4}{t}=1+\frac{\alpha}{\sqrt{3N}}
\ee
We end up with:
\be
\lim_{N\rightarrow\infty}
\left(\frac{4N}{3}\right)^{\frac{1}{4}}Z_N(y,t) =
\int_{-\infty}^{\infty} du e^{-u^4-\alpha u^2+\xi u }
\equiv \zeta(\xi,\alpha)
\ee
The above equation explicitly shows that
keeping $\xi$ and $\alpha$ fixed, 
while taking $N$ to infinity will make the function 
$\left(\frac{4N}{3}\right)^{\frac{1}{4}}
Z_N(y,t)$ converge point-wise to the $\alpha$- and $\zeta$- dependent limit given by $\zeta(\xi,\alpha)$.
Looking at equation~(\ref{serieseq}), we see that corrections will go as
powers of $\frac{1}{\sqrt{N}}$. 
A plot of the logarithmic derivative of $\zeta$ with respect to
$\xi$ in Figure~\ref{smoothfig}.  
shows that the double scaling limit
provides a smoothed version for the non-analyticity 
discussed in (\ref{heur}).

\section{Formulation of the large $N$ universality hypothesis in dimensions 
2,3,4. }
We now abstract from the two dimensional case a hypothesis expected to hold
also for Euclidean $SU(N)$ gauge theory in dimensions 3 and 4. 
We first formulate the statement in continuum ignoring renormalization,
and next provide a precise formulation using lattice gauge theory as a 
constructive definition of continuum YM.

\subsection{Continuum formulation -- ignoring renormalization.}

Suppose we have a Wilson loop associated with a curve ${\cal C}$, $W({\cal C})$. 
Suppose the loop ${\cal C}$, is parametrically described by a closed, 
non-self-intersecting curve $x_\mu (s), s \in [0,1]$. This description is 
redundant under re-parameterizations of the curve.  
Consider this curve together with an infinite family of scaled versions of
it: ${\cal C}(m)$, described parametrically by $x_\mu(s,m)=\frac{1}{m} x_\mu 
(s)$,with $m > 0.$ If we collect all these families we obtain the space of all 
loops. We wish to think about a single loop ${\cal C}(m)$ 
as being labeled by its shape ${\cal C}(*)$, 
which is the label
of its scaled family and is described by dimensionless
parameters, and a particular scale $m$ which identifies it
uniquely within the family and is of dimension mass.  
We now pick a loop shape and look at the family of operators $W(m,{\cal C}(*))=
W({\cal C}(m))$. We are interested in the behavior of $W(m,{\cal C}(*))$ 
as we vary $m$, keeping ${\cal C}(*)$ fixed. More specifically, we are looking 
at \be
O_N (y,m,{\cal C}(*))=\la \det (e^{\frac{y}{2}}+e^{-\frac{y}{2}} 
W(m,{\cal C}(*))\ra
\ee
with particular interest focused on the region where $y$ is close to $0$.

The first part of the hypothesis is that the definition makes sense, meaning
that $O_N(y,m,{\cal C}(*))$ is well defined, and that indeed there exists some 
scale $m_c$ of the basic loop shape ${\cal C}(*)$
at which the Wilson matrix 
undergoes the DO large $N$ phase transition. The part of the hypothesis that 
has to do with large $N$ universality says that there exists a (non-universal)
normalization
${\cal N}(N,m,{\cal C}(*))$, dependent on $N$, $m$ and the loop shape, 
and
finite dimensionless parameters $a_1({\cal C}(*)),a_2({\cal C}(*))$ such that 
\be
\lim_{N\rightarrow\infty} {\cal N}(N,b,{\cal C}(*))
O_N\left(y=
\left(\frac{4}{3N^3}\right)^{\frac{1}{4}}\frac{\xi}{a_1({\cal C}(*))},
m=m_c\left [1+\frac{\alpha}{\sqrt{3N}a_2({\cal C}(*))}\right ]\right) = 
\zeta(\xi,\alpha)\ee

\subsubsection{Two dimensions: no renormalization needed.}

In two dimensions, for a non-self-intersecting loop, the dependence on ${\cal 
C}$ comes only through its total enclosed 
area; there is no dependence on ${\cal C}(*)$, the loop
shape, but only on $m$, its scale, which can be defined as the square root of
the inverse of the area. Two dimensional YM has a dimensional coupling which 
does not renormalize and simply keeps track of dimensions. The issue of
renormalization does not arise at all.  
We may as well regard 
the area as dimensionless and set the coupling constant to unity. The 
dimensionless positive parameter $t$ of the random matrix model corresponds to 
this dimensionless area. It is convenient to change notation, from $t$ to $b$,
\be
b=\frac{4}{t}
\ee
and view $b$ as $m^2$ in our discussion above.  
$m_c=1$, since $t_c=4$. When $m$ increases the loop shrinks.  

We can now summarize our previous findings in two dimensions as follows:
Consider
\be
{\tilde O}_N(y,b)=
\left(\frac{N}{12}\right)^{\frac{1}{4}} \sqrt{\frac{2\pi}{Nb}}
\frac{e^{\frac{N}{2b}}}{2^N} \left < \det \left( e^{\frac{y}{2}}
+e^{-\frac{y}{2}} \prod_{i=1}^N U_i \right) \right>
\label{modeleqn}
\ee
${\tilde O}$ is proportional to $O$ but the normalization is
$b$ dependent. The large $N$ universal content  is independent of the 
prefactor, so long as the normalization is smooth in $b$ at the point 
$b=b_c$; therefore the difference between ${\tilde O}$ and $O$ is immaterial. Using (\ref{znyt}), we can see that
\be
{\tilde O}_N(y,b)=\left(\frac{N}{12}\right)^{\frac{1}{4}}
\int d\rho e^{N\left[\ln\cosh\rho-\frac{b}{8}
(2\rho-y)^2\right]}. 
\label{mock2d}
\ee
Defining 
$\xi$ and $\alpha$ by
\be
y=\left(\frac{4}{3N^3}\right)^{\frac{1}{4}}\xi;\ \ \ 
b=1+\frac{1}{\sqrt{3N}}\alpha
\label{yarho}
\ee
and expanding in $\frac{1}{\sqrt{N}}$, we obtained:
\be
\lim_{N\rightarrow\infty} {\tilde O}_N(y,b)=
\zeta(\xi,\alpha)=\int du e^{-u^4 -\alpha u^2 + \xi u}
\label{airy}
\ee

This is how the universality hypothesis is realized in two dimensions, by 
construction. 

\subsection{Lattice formulation -- completely defined.}

Several of the choices we shall make are not conceptually essential, but
they help streamline the discussion.  

\subsubsection{Shape and scale of curves on the lattice.}

We start by replacing space-time by a hypercubic lattice in $d$ dimensions.  
This hypercubic lattice will be viewed as dimensionless, a collection of
vertices, or sites, labeled by $x_\mu \in Z, \mu=1,..,d$, and the 
shortest arcs, or links, connecting them. One adds an orientation to the
links: this means that a link parallel to the $\mu$-axis 
$\mu=1,...,d$ can be traversed 
in the direction of its orientation $(+\mu)$, or in the opposite sense $(-\mu)$. 
This setup is used to define approximations
to curve shapes. The curve shape is replaced by a contiguous sequence of
links, where the angles between any two links have to be a multiple of 
ninety degrees. Symbolically, the curve is represented by an ordered
sequence, $(\mu_1,\mu_2,...,\mu_L)$ where $\mu_i=-d,-(d-1),....,d-1,d$. 
The curve is closed when $\sum_{i=1}^L \delta_{\nu,\mu_i} =0$ for 
$\nu=1,2,..,d$. When the curve is closed there is a redundancy under cyclic 
shifts of the sequence.  
The curve is non-self-intersecting if every site is visited no more than once.  
The total number of links, $L$, determines how good the approximation
is. In the continuum limit one needs to take $L$ to infinity.  

A scale parameter is attached to the curve shape by the ``dynamics''.  
To each link we attach an $SU(N)$ unitary matrix $U$. There is a joint
probability distribution for all link matrices $U$, which is parameterized
by a positive parameter that we again call $b$. The mass scale $m$ is determined 
by $b$ and the relationship is monotonic: $m(b)\to 0$ as $b\to\infty$. 
The continuum limit is obtained 
by taking $L\to\infty$, $b\to \infty$, in such a way that $m(b) L=l$ stay 
finite. One can then arbitrarily introduce a unit of length to give $l$ 
engineering dimensions.  

For a fixed continuum curve, its shape ${\cal C}(*)$ is obtained from the 
lattice sequence in the limit when $L\to\infty$. Simultaneously with that limit 
one needs to take $b\to\infty$, while the product $m(b) L=l$ stays finite. 
$l$ determines the scale of the curve ${\cal C}$, and plays the role of 
the parameter $\frac{1}{m}$ in the continuum discussion. One way to investigate
what happens as the continuum scale $m$ goes through its critical value
for a given curve shape, is to vary $b$ at a fixed lattice curve with a fixed 
$L$. The universality hypothesis makes a prediction about this behavior; this 
prediction is approximate in that the parameters $a_1, a_2$ are $L$ dependent 
but becomes accurate as $L\to\infty$. The order of the
limits $L\to\infty$ and $N\to\infty$ is assumed to not matter, although there
are some limitations on the ranges.  

\subsubsection{Regularization of perimeter and corner divergences.}

To make the prediction of universality quantitative we need to assure that the
lattice version of $O_N(y,b)$ is well defined and has a finite continuum limit. We need to eliminate
corner and perimeter divergences.
They are eliminated by replacing the link 
matrices $U$ in the standard definition of $W$ 
by smeared versions, denoted by $U^{(n)}$, where 
$n$ is an integer.  

We employ APE smearing \cite{ape10}, defined
recursively from $n=0$, where the smeared matrix is equal to
$U_\mu(x)$. Let $\Sigma_{U^{(n)}_\mu (x;f)}$ denote the ``staple''
associated with the link $U^{(n)}_\mu(x;f)$ in terms of the entire
set of $U^{(n)}_\nu(y;f)$ matrices. One step in the recursion
takes one from a set $U^{(n)}_\mu (x;f)$ to a set $U^{(n+1)}_\mu
(x;f)$:
\begin{eqnarray}
X^{(n+1)}_\mu (x;f)= (1-|f|) U^{(n)}_\mu (x;f)+\frac{f}{2(d-1)}
\Sigma_{U^{(n)}_\mu (x;f)}\nonumber\\ 
Y^{(n+1)}_\mu (x; f
)=X^{(n+1)}_\mu (x;f) \frac{1}{\sqrt{[X^{(n+1)}_\mu (x;f)]^\dagger
X^{(n+1)}_\mu (x;f)}}\nonumber\\
U^{(n+1)}_\mu(x;f)=\frac{Y^{(n+1)}_\mu (x; f)}
{\det^{\frac{1}{N}}\left[Y^{(n+1)}_\mu (x; f)\right]}
\end{eqnarray}
In the simulation, one never encounters a situation where the
unitary projection in the above equations stalls because $X^{(n)}$
is singular. In other words, smearing is well defined with
probability one. 

$U^{(n)}_\mu (x;f)$ transforms under gauge transformations the
same way as $U_\mu (x)$ does. For definiteness we restrict our 
subsequent discussion to rectangular loops of sides $L_1$ and $L_2$
which fit into a two dimensional plane in the $d$ dimensional
Euclidean space time.

Our smeared Wilson loop operators,
$\hat W [L_1,l_2 ; f ; n]$ are defined as ordered products around
the $L_1 \times L_2$ rectangle restricted to a plane. $L_\alpha$
are integers and give the size of the loop in units of the lattice
spacing. When the traversed link starts at site
$x=(x_1,x_2,...x_d)$ $ x_\mu\in Z$ and connects to the
neighboring site in the positive direction $\mu$, $x+\mu$, the
link matrix is $U^{(n)}(x;f)$, while when this oriented link is
traversed in the opposite direction, the link matrix is
$U^{\dagger (n)}(x;f)$. $\hat W$ depends on the place the loop was
opened, but its eigenvalues do not. The set of eigenvalues is
gauge invariant under the fundamental gauge transformation
operating on $U_\mu (x)$.

We adjust the parameter
dependence in $\hat W$ such that the $N=\infty$ transition points
which are seen to occur on the lattice, survive in
the continuum limit in which the lattice coupling $b$ is taken to
infinity together with $L_{1,2}$ in such a way that the physical
lengths $l_\alpha = L_\alpha m(b)$ are kept fixed. (
$l_1/l_2=L_1/L_2$ is independent of $b$, and represents the loop
shape; our previously defined scale $l$ is $l=2m(b)(L_1+L_2)$). 

We set the number of smearing steps $n$ to be proportional to the
perimeter square (we restricted the loop sizes to even
$L_1+L_2$), $n=\frac{(L_1+L_2)^2}{4}$. The physical
sizes of the loop are $l_\alpha$. We have set 
$n=\frac{(L_1 + L_2 )^2}{4}$ because in physical terms
the product $f n$ is a length squared. This is
because smearing is a random walk that
fattens the loop and the thickness grows as the square root of the
number of smearing steps. Our choice for $n$ makes $f$ a
dimensionless parameter in the physical sense; on the lattice $f$
is actually bounded to an interval of order one.  
The effect of smearing is easy to understand in perturbation
theory where one supposes that each individual step in the
smearing iteration can be linearized. Writing $U^{(n)}_\mu
(x;f)=\exp(iA^{(n)}_\mu (x;f))$, and expanding in $A_\mu$ one
finds~\cite{pertsmear11}, in lattice Fourier space:
\be
A^{(n+1)}_\mu (q;f)= \sum_\nu h_{\mu\nu} (q) A^{(n)}_\nu (q;f) 
\ee
with
\be
h_{\mu\nu} (q)= f(q)(\delta_{\mu\nu} - \frac {{\tilde q}_\mu
{\tilde q}_\nu}{\tilde q^2})+\frac {{\tilde q}_\mu {\tilde
q}_\nu}{\tilde q^2} \ee where $\tilde q_\mu = 2\sin
(\frac{q_\mu}{2})$ and
\be
f(q)=1-\frac{f}{2(d-1)}\tilde q^2
\ee
The iteration is solved by
replacing $f(q)$ by $f^n (q)$, where, for small enough $f$,
\be
f^n(q) \sim e^{-\frac{f n}{2(d-1)} \tilde q^2}
\ee

Much larger loops should not be smeared with an $f n$ factor that
keeps on growing as the perimeter squared; rather, for a square
loop of side $L$, for example, the following choice would be
appropriate:
\be
f=\frac{\tilde f}{1+M^2 L^2}
\ee
Here, $M$ is in lattice units, $M=\Gamma m(b)$. $\Gamma$ is a hadronic
scale chosen so that at the large $N$ transition, $\Gamma l$ is
less then $0.01$, say. 

The new parameter $f$ should be considered as fixed once and for all; its
exact value is unimportant so long as it is reasonable. However, if that
value is changed by some modest amount the critical loop size will change too.
This critical loop size is non-universal; only the fact that such a critical
value exists within some reasonable hadronic range is universal.  

Smearing provides a means to regularize the basic observable and allows us to
proceed finally to the lattice formulation of the large $N$ universality
hypothesis. For simplicity, we formulate it only for square Wilson loops,
denoting by $W$ the operator constructed from smeared link variables.

\subsubsection{Universality hypothesis for square lattice Wilson loops.}

We assume to be given a table (the data) 
with numerical values for the expectation value of
\be
O_N(y,b)=
\left< \det (e^{\frac{y}{2}}+e^{-\frac{y}{2}}W)\right>
\ee
for an $L\times L$ Wilson loop at an inverse 
't Hooft rescaled gauge coupling $b$. The hypothesis says that $O_N(y,b)$ will 
exhibit critical behavior at $b=b_c(L)$ and $y=0$
as $N\rightarrow\infty$. There, it will obey 
large $N$ universality, which means that there exists a ${\cal N}(b,N)$, 
smooth in $b$ at $b=b_c$, such that: 
\be
\lim_{N\rightarrow\infty} {\cal N}(b,N)
O_N\left(y=
\left(\frac{4}{3N^3}\right)^{\frac{1}{4}}\frac{\xi}{a_1(L)},
b=b_c(L)\left [ 1 +\frac{\alpha}{\sqrt{3N}a_2(L)}\right ] \right) = 
\zeta(\xi,\alpha)\ee
${\cal N}(b,N)$ is a normalization factor 
similar to the one in (\ref{modeleqn}).  

\subsubsection{Large $N$ universality holds already before the continuum limit.}

Even at finite $L \ge L_0$, where $L_0$ is some finite number there
will be a large $N$ phase transition in loops. Our hypothesis includes
the belief that this transition will be in the DO universality class
even before the continuum limit is taken. Thus, for simple enough
loops it always makes sense to define $b_c(L), a_1(L), a_2(L)$.
If all three parameters approach the continuum limit in the
standard manner, then 
large $N$ universality is a property of the continuum
limit.

This is somewhat similar to spontaneous chiral symmetry on the
lattice.  Using the overlap action for example~\cite{ovlap}, we can
define a pion decay constant at finite lattice spacing, by
relating the pion mass for small bare quark masses
using standard chiral symmetry considerations. That all
this survives the continuum limit amounts simply to checking that
physical quantities have smooth limits, 
approached in standard ways.  The
key is to have a good lattice definition that preserves the essential
ingredient of the phenomenon. When there is no reason, the continuum
limits and other critical behaviors do not interfere with each
other. However, if the lattice regularization is faulty, for example
ignoring perimeter effects in the case of Wilson loops, or choosing a
Wilson type of fermionic action in the chiral case, one will have
interference with the continuum limit. This is not to say that these
problems cannot be overcome --only the analysis becomes more murky and
delicate.

\subsubsection{How to test for universal large $N$ behavior numerically?}

We test for large $N$ universality hypothesis as follows:
Obtain 
estimates for $b_c(L)$, $a_1(L)$, 
$a_2(L)$ denoted by $b_c(L,N)$, $a_1(L,N)$ and
$a_2(L,N)$ 
from data at various values of $N$ 
assuming $N$ is already large enough to use the asymptotic formulas. 
\begin{itemize}
\item Show that all three $N$-dependent quantities have a well
defined limit as $N\to\infty$, which is approached in a way consistent with
large $N$ universality.  
\item Show that $b_c(L)$, $a_1(L)$, 
$a_2(L)$ have finite limits as $L\to\infty$ (which are correlated with 
taking $b\to\infty$ keeping the physical loop size fixed). Moreover, 
these limits should be approached in the manner expected 
of normal physical field theoretical observables (that is the sub-leading 
corrections can be organized by dimensional analysis restricted by symmetry 
considerations).
\end{itemize}

\subsubsection{The estimates $b_c(L,N)$ for $b_c(L)$.}\label{bcest}

$O_N(y,b)$ is an even
function of $y$ because $W\in SU(N)$. It is obvious from (\ref{airy})
that $\zeta(\xi,\alpha)$ is an even function of $\xi$. 
Let
\be
O_N(y,b)= C_0(b,N) + C_1(b,N) y^2 +  C_2(b,N) y^4 + \cdots
\label{taylor}
\ee
be the Talyor's series for $O_N(y,b)$. 
At some fixed value of $L$ we consider the following quantity,
derived from the average characteristic polynomial of the regularized Wilson
loop:
\be
\Omega(b,N) = \frac{ C_0(b,N) C_2(b,N)}{C_1^2(b,N)}. 
\ee
$\Omega(b,N)$ is essentially a ``Binder cumulant''~\cite{binder}.  
The normalization ${\cal N}(b,N)$ and any rescaling of $y$ drop 
out from $\Omega (b,N)$. Therefore, if $N$ is large enough, 
and if we set $b=b_c(L,N=\infty)$ we should get a value close to the number
$\Omega (b_c,\infty)$. We define an approximation to $b_c(L,N=\infty)$, 
$b_c(L,N)$, by the equation:
\be
\Omega(b_c(L,N),N) = 
\frac{\Gamma(\frac{5}{4}) \Gamma(\frac{1}{4})}{6 \Gamma^2(\frac{3}{4})} = 
\frac{\Gamma^4(\frac{1}{4})}{48\pi^2}= 0.364739936
\label{method1}
\ee
Viewing the $u$ integrand in 
(\ref{airy}) as performing an average over $u$ dependent observables, 
we would write, $C_k=\frac{\la u^{2k}\ra}{(2k)!}$. For $|\alpha|>>1$ 
we can assume that $u$ is
approximately distributed as a Gaussian. For $\alpha >0$ the mean is zero, 
$\la u\ra =0$, and using $\frac{\la u^4 \ra}{{\la u^2\ra}^2}=3$ gives 
$\Omega =\frac{1}{2}$. If $\alpha <0$ 
there are two nonzero saddles of the same absolute magnitude 
$\la u\ra\ne 0$; these saddles dominate over fluctuations, giving 
$\Omega =\frac{1}{6}$.  The full function $\Omega(\alpha)$ is shown 
in Figure~\ref{omegas}. 

At $\alpha=0$ $\Omega$ comes out pretty close to the arithmetic 
average of the two asymptotic values: $\frac{1}{2}(\frac{1}{2}+\frac{1}{6})$.  
The exact number is given in (\ref{method1}); it was obtained from (\ref{airy}) 
using 
\be
\int_{-\infty}^\infty du u^{2k} e^{-u^4} =\frac{1}{2}\Gamma
\left ( \frac{2k+1}{4} \right )
\ee

Expanding ${\tilde O}(y,b)$ in (\ref{mock2d}) to order $y^4$ leads to 
explicit expression for $\Omega(b,N)$.
Figure \ref{omegan} shows $\Omega(b,N)$  for different 
values of $N$.   
In two dimensions the exact $\Omega (b,N)$ connects
monotonically the two extremes, $1/6$ and $1/2$ as $b$ varies
from far below $b_c$ to far above $b_c$.  
If one takes $N\to\infty$, there is a discontinuous jump at $b=b_c$, 
between the above two asymptotic values. The double scaling 
limit of $\Omega(b,N)$, which produced $\Omega(\alpha)$, smoothed 
out the jump but maintained the asymptotic behavior of the exact 
expression at finite $N$.

From the data we get estimates of $C_i(b,N)$, $i=0,1,2$,
from which we extract $\Omega(b,N)$. We then use (\ref{method1})
to obtain an estimate of $b_c(L,N)$. 

$b_c(L,N)$ has been constructed from the free energy of the combined 
system of gauge fields and fermions used to represent
the characteristic polynomial. Therefore, ordinary $N$-power
counting rules should apply, and we expect $b_c(L,N)$ to approach
$b_c(L,\infty)\equiv b_c(L)$ as a series in $\frac{1}{N}$.  

\subsubsection {The estimates $a_2(L,N)$ for $a_2(L)$.}\label{a2est}

The parameter $a_2(L,N)$ is
obtained by first setting
\be
b=b_c(L,N)\left[ 1 + \frac{\alpha}{\sqrt{3N}a_2(L,N)}\right],
\label{bscale}
\ee
where $b_c(L,N)$ has been defined above.  Next we write 
the derivative of $\Omega$ with
respect to $\alpha$ and set the result equal to the corresponding 
universal number in the large $N$ limit.  
\be 
\left.\frac{d\Omega(b,N)}{d\alpha} \right |_{\alpha=0} =
\left.\frac{1}{a_2(L,N)\sqrt{3N}} \frac{d\Omega}{db}\right |_{b=b_c(L,N)}
= \frac{\Gamma^2(\frac{1}{4})}{6\sqrt{2}\pi}
\left( \frac{\Gamma^4(\frac{1}{4})}{16\pi^2} -1\right)
=0.0464609668
\label{a2ex}
\ee
$\frac{d\Omega}{db}$ would be close to maximal at $b=b_c$; hence 
$\frac{d\Omega}{db}$ varies relatively little as $b$ stays close to $b_c$. Since 
$b_c$ is not known to infinite accuracy the reduced sensitivity on the exact 
value of $b_c$ is an advantage which motivates this choice for defining 
$a_2(L,N)$.   
Unlike $b_c(L,N)$, the definition of $a_2(L,N)$ involves 
going into the large $N$ critical regime around $b_c(L,\infty)$ and
non-standard powers of $N$ come in.  
Taking this into account, we expect $a_2(L,N)$ to approach 
$a_2(L,\infty)\equiv a_2(L)$ as a power series in $\frac{1}{\sqrt{N}}$. 

\subsubsection {The estimates $a_1(L,N)$ for $a_1(L)$.}\label{a1est}

We substitute 
\be
y=
\left(\frac{4}{3N^3}\right)^{\frac{1}{4}}\frac{\xi}{a_1(L,N)}
\ee
in (\ref{taylor}). We then form a ratio whose value at infinite $N$ is again
a universal number we can easily compute. 
\be
\sqrt{\frac{4}{3N^3}} \frac{1}{a_1^2(L,N)}
\frac{C_1(b_c(L,N),N)}{C_0(b_c(L,N),N)}
= \frac{\pi}{\sqrt{2}\Gamma^2(\frac{1}{4})}=0.16899456
\label{a1ex}
\ee
This relation defines $a_1(L,N)$. 

Similarly to $a_2(L,N)$, the definition of $a_1(L,N)$ involves 
going into the large $N$ critical regime around $b_c(L,\infty)$. 
Consequently, we expect $a_1(L,N)$ to also approach 
$a_1(L,\infty)\equiv a_1(L)$ as a power series in $\frac{1}{\sqrt{N}}$.

\subsection {Example of a universality test on synthetic two dimensional data.}

In two dimensions we work already in the limit $L=\infty$. Our main objective is 
to check what kind of finite $N$ data could be used to produce the known 
infinite $N$ values of $a_1,a_2,b_c$, using the above procedures (with $L$  
eliminated) to define $a_1(N), a_2(N), b_c(N)$. The extrapolation to infinite 
$N$ is done using a series in $\frac{1}{N}$ for $b_c(N)$ and a series in 
$\frac{1}{\sqrt{N}}$ for $a_1(N), a_2(N)$ as explained above.  
The values of $N$ in Figure~\ref{omegan}  
were chosen to match the ones employed 
in the three dimensional simulation.

$\Omega(b,N)$, as a continuous function of $b$ for a fixed $N$,
defines via (\ref{method1}) the number $b_c(N)$.  
With $b_c(N)$ so determined, we use (\ref{a2ex}) 
to determine $a_2(N)$ from $\frac{d\Omega(b,N)}{db}\left. \right |_{b=b_c(N)}$.  
Further, we use (\ref{a1ex}) to find the value of $a_1(N)$ from 
$\frac{C_1(b_c(N),N)}{C_0(b_c(N),N)}$. Figures \ref{bcn},\ref{a2n},\ref{a1n}
show what can be done with ``perfect'' data 
for $N=17,23,29,37,41,47$. 
The $N\to\infty$ estimate of $b_c$ is the most accurate followed
by the estimates of $a_1$ and $a_2$. 
This is typical in that we expect (and need) an accurate estimate of the
critical point while the estimate of the amplitudes come at lower accuracy.  

While the synthetic data was produced only at values of $N$ that are practical
also in three dimensions, it 
has three features that are not in common with
lattice data obtained by Monte Carlo simulations:
\begin{enumerate}
\item There are no statistical errors. 
\item We know $\Omega(b,N)$
and $\frac{C_1(b_c(N),N)}{C_0(b_c(N),N)}$
as continuous functions of $b$. The numerical
simulation will be performed only on a discrete set of 
$b$ values that brackets $b_c(N)$
and one will need to interpolate. 
\item We know $\frac{d\Omega(b,N)}{db}$ exactly.  
A direct numerical estimate of this derivative would involve linear
combinations of connected correlations of $C_i(b,N)$ with
the plaquette operator. This has large statistical
errors and is expected to be too expensive
to compute accurately. Therefore, we shall not have a direct
numerical estimate of the derivative and will extract it
from the interpolation of $\Omega(b,N)$ we have already used when 
determining $b_c (N)$.   
\end{enumerate}
The synthetic data is used to 
indicate to us what ranges of $N$ are needed 
to reliably extrapolate the three parameters to their 
$N\to\infty$ limits.   
The conclusion is that it is possible to carry out quite accurate
estimates of 
$b_c$ and reasonably accurate estimates of
$a_2$ and $a_1$ in that $N\to\infty$ limit
from data obtained
at values of $N$ which are within the range of Monte Carlo simulations
in dimensions higher than two. However, the differences we have listed above 
are sources of extra systematic and stochastic errors that we shall need to 
control.

\subsection{Volume dependence and large $N$ reduction.}

In a precise sense, YM theory in 3 or 4 dimensions on a finite torus
becomes independent of torus size at infinite $N$ if
the torus is larger than some critical torus~\cite{largenred}.  
The size has to be large enough
for the system to be in the so called 0c phase at infinite $N$.  
In 0c, traces of Wilson loops in representations of finite dimension are equal
to their infinite volume values up to corrections of order $\frac{1}{N^2}$.  
0c is characterized by all Polyakov loops having uniform eigenvalue 
distributions.  
Using the fermionic representation of the average characteristic polynomial 
we expect that 
\be
\frac{1}{N}\log(\la\det(z-W)\ra)\ee
will also be independent of the volume at leading order at large $N$, with 
corrections going as $\frac{1}{N}$. Looking at the powers of $N$ that enter into 
the function $\zeta(\xi,\alpha)$ we conclude that it also should be independent 
of the volume. However, one expects the sub-leading 
corrections in $\frac{1}{N}$, 
which are volume dependent (and non-universal even at 
infinite volume) to be relatively larger than for traces of Wilson loops.  
The reason to expect slower convergence to the infinite $N$ limit is
that $\zeta(\xi,\alpha)$ describes a large $N$ critical regime, where taking 
enough derivatives with respect to some parameter would produce quantities that 
diverge in the ordinary (without double scaling) large $N$ limit. Obviously, 
nothing is supposed to diverge at finite $N$, so sub-leading corrections must be 
large. These sub-leading corrections will be even more significant at
smaller
volumes. 

Thus, although large $N$ reduction can be exploited, one needs to carry out an 
explicit check to determine how much contamination of the final estimates has 
been caused by using relatively small volumes.  

\section{Three dimensions.}

We use an ordinary single plaquette Wilson action defined on a
hypercubic lattice. 
Our simulation method
employs a combination of heat-bath and over-relaxation updates and
``thermal equilibrium'' is achieved in reasonable lengths of computer time. 
We keep our statistical errors small relative to systematical ones.  
Throughout, we use $b$ for the lattice gauge coupling which is the
inverse bare 't Hooft coupling. It is related to the conventional
lattice coupling $\beta$ by
\be
b=\frac{\beta}{2N^2}=\frac{1}{g_{\rm YM}^2N}\ee
and $b$ already has the right power of $N$
extracted.
It is useful to consider the tadpole improved coupling,
$b_I=be(b)$ where $e(b)$ is the average value of the
trace of plaquette operator. To facilitate a translation
from $b$ to $b_I$, we have plotted $e(b)$ 
in Figure \ref{plaq} for the range
of $b$ used in this paper. 

We started the project by carrying out preliminary simulations, intended to 
identify a convenient value for the
parameter $f$. We established, in a way 
similar to
our earlier work in four dimensions~\cite{ourjhep}, 
that as $L_1=L_2=L, f, n$ are varied, at specific
lattice couplings, the spectrum of $\hat W [L_1=L_2=L; f ; n] $
opens a gap for very small and/or very smeared loops. This gap
closes for very large and/or very lightly smeared loops. 
We worked\footnote{We would like to thank Alejandro de la Puente
for some preliminary work in this direction.} at $N=37$
on a $8^3$ lattice. Keeping $b$ fixed, we varied $f$ and obtained
an estimate of the gap using the technique described in~\cite{ourjhep}.  
This was done for several  Wilson loops of size $L^2$, $L$  
ranging from $L=2$ to $L=10$. In this manner, we obtained
estimates for $f_c(L,b)$, the critical value of $f$ at which the 
gap opens around eigenvalue $-1$  
when the smearing of $L\times L$ Wilson loops
is steadily increased at fixed $b$. The function $f_c(L,b)$ has a continuum 
limit, obtained when $L$ and $b$ go to infinity in the usual correlated way. 
This was tested employing five different values of 
coupling, $b=0.85,0.9,1.0,1.1,1.2,1.3$; we made sure that all these 
couplings are in the $0$c phase~\cite{threed} for our $8^3$ lattice.  
We found that all the values $f_c(L,b)$ fall on a common curve 
when plotted as a function $L/b_I (b)$ as shown in Figure \ref{fcrit}.  

Based on this work, we chose to carry out the more detailed 
analysis of the large $N$ critical region, which is the main topic
of this paper, at $f=0.03$. Other values of $f$, between $0.02$ and $0.04$, 
might have served as well, although many numbers, including $b_c$ and $a_2$, 
would have changed by modest amounts.  Much higher values of  
$f$ are counter indicated at this stage of our research 
because we want to avoid finite volume effects and 
therefore wish to keep $L$ below $8$. A lattice of size $8^3$ affords 
reasonably speedy simulation, even at $N=47$, but the cost
quickly escalates when the lattice size is increased.  A more detailed 
discussion of finite size effects will be presented below. 

\subsection{Details of the numerical analysis.}

Our simulations are carried out for prime numbers for $N$ to ensure that 
the phase 0c does not decay into phases related to proper 
subgroups of $Z(N)$.  This is a precaution; it is possible that one
could also work with non-prime values of $N$. 
We employed six different values of $N$,
namely $N=17,23,29,37,41,47$. There are three more primes in this
range, but they are so close to other primes, that we did not expect the
extra information to be worth the effort. 
In order to check for volume dependence 
we obtained data for $2^2$ Wilson loops on $3^3$, $4^3$ and $6^3$ lattices 
and for the $3^2$ loop on $4^3$ and $8^3$ lattices.  For our main study of the 
double scaling limit we produced data for loops of larger sizes, $4^2$,
$5^2$ and $6^2$, all on a single lattice size, $8^3$. For each square 
loop $L^2$, and for each value of $N$ we carried out a series of simulations in  
a range of $b$ separated by small steps $\Delta b$.  
Table~\ref{tab1}, which provides the intermediate numerical output used in 
the study of the double scaling limit, also lists all values of $L$ and $N$ 
along with the lattice volume $V$.  

After equilibration, for which we typically allowed several thousands
of lattice passes, the different steps were separated by 1000
passes. We tested the autocorrelation for our observable and saw that
we exceeded it by enough not to have to worry about the independence
of our samples. For each entry in the Table~\ref{tab1} we did
somewhere between $31$ to $48$ separate simulations on parallel nodes
in one of our PC clusters. Measurements on a single Wilson loop
was averaged over the whole lattice for all orientations. 
Statistical errors obtained from the measurements on
several configurations were always estimated
by jackknife with single elimination.

In each run we collected data for 30 values of $y$ around zero, at equally 
spaced points, where the range was determined to be fixed in terms of the 
corresponding rescaled $\xi$ variable, assuming $a_1=1$ at all $N$, $L$ and $b$: 
$0 \le \xi \le 3$. There is no need to collect data also at negative values of 
$\xi$, since the symmetry under a sign flip of $y$ is exact.  

In order to perform a cross-check of our procedure described in
(\ref{bcest}), (\ref{a2est}) and (\ref{a1est}),
the first type of data we collected is for the observable $O(y,b,L)$:
\be
O(y,b,L)=\la \det\left ( e^{\frac{y}{2}}+ e^{-\frac{y}{2}} W(b,L)\right )\ra
\ee
More specifically, we collected data for its logarithmic derivative with
respect to $y$ directly; this means that at fixed $y,b,L$ 
for each gauge configuration and for each loop one keeps two numbers,
the determinant and its derivative with respect to $y$. 
These numbers are summed over all translations of the loop and these two
numbers are stored for subsequent gauge averaging when the analysis is done.  
For a fixed $N$ and $L$, the data makes up a two dimensional rectangular grid
in the $\xi,\alpha$ plane.  

We used a nonlinear fitting 
routine to find a best match of the logarithmic derivative with respect to $y$ 
of $O$ to the logarithmic derivative with respect to $\xi$ of 
the double scaling function $\zeta(\xi,\alpha)$. This produces three parameters
$b_c(L,N),a_1(L,N),a_2(L,N)$ which can be extrapolated later on, first in $N$,
and subsequently in $L$. The fitting routine we used was based on the 
Levenberg-Marquart method and the implementation in~\cite{numrec}. 
The logarithmic derivative with respect to $\xi$ of 
the double scaling function $\zeta(\xi,\alpha)$ was calculated using gaussian
integration over several intervals to high accuracy. 
In addition to producing estimates to the parameters as mentioned, this showed 
us that indeed one approaches the double scaling limit. We first tested the 
nonlinear fitting method on synthetic data in two dimensions as reported 
in~\cite{pospar}. This will not be reviewed here again.  
As a method of estimating
parameters, the simultaneous nonlinear fit has the drawback that all parameters
now have corrections of the order $\frac{1}{\sqrt{N}}$. As we have seen, for
$a_1(L,N),a_2(b,N)$ this is unavoidable, but for $b_c(L,N)$ we can do better. 
The nonlinear simultaneous fit mixes the corrections up and therefore is not
the best way to prepare the ground for the large $N$ extrapolation.  

The second type of data we collected is used for determining the parameters from 
the behavior around $y=0$ that are expressed by the three coefficients 
$C_i(b,N,L)$.  

We first obtain an estimate for $\Omega(b,N,L)$.  
Figure \ref{omega47} shows a sample plot of $\Omega(b,N,L)$ as a function
of $b$ for $N=47$ and $L=3$ on a $4^3$ lattice. 
The behavior is similar to the two dimensional case. 
The top and bottom horizontal lines are the limits at
weak and strong coupling, $1/2,1/6$ respectively. The middle line
is the expected value at critical coupling in the $N\to\infty$
limit as given by the right hand side of (\ref{method1}). 

We focus on a region of $\Omega(b,N,L)$ that is bounded by the two
horizontal lines that are on either side of the middle line in
Figure \ref{omega47}. We view $b$ as a function of
$z=\Omega(b,N,L)-0.364739936$ in this region and
use a linear three-parameter fit to a degree 2 polynomial:
\be
b=b_c(L,N) + \frac{1}{\frac{d\Omega}{db}|_{b=b_c(L,N)}} z
+\beta z^2
\ee
This gives us our determination for $b_c(L,N)$. 
With the help of this same polynomial we then extract $a_2(L,N)$, using 
(\ref{a2ex}).  

Next, we analyze the numbers for the ratio $\frac{C_1(b,N,L)}{C_0(b,N,L)}$ as 
follows: We take $\frac{C_1(b,N,L)}{C_0(b,N,L)}$ as a 
function of $z$ in the same region we used above 
and again carry out a linear three-parameter fit to a degree 2 polynomial.  
$\frac{C_1(b_c(L,N),N,L)}{C_0(b_c(L,N),N,L)}$ is set as the leading  coefficient 
 in this fit.  Finally, $a_1(L,N)$ is extracted using equation (\ref{a1ex}).  

\subsection{Extrapolation to infinite $N$.}

We take the $6^2$ loop on $8^3$ lattice as a sample case and plot the
results from the linear fit using described in (\ref{bcest}),
(\ref{a2est}) and (\ref{a1est}).  The solid circles in figures
\ref{bc6},\ref{a26} and \ref{a16} show the results for $b_c(L,N)$,
$a_2(L,N)$ and $a_1(L,N)$ respectively.  The extrapolation to infinite
$N$ was done using a three term series. One cannot use smaller $N$
value when doing this and larger $N$ values are two expensive in
computer time to produce, the simulation time growing as $N^3$. Some
systematic errors are induced by this extrapolation; one can get a
feel for it by using more, or less, powers of $N$ in the series.
The open circles in figures \ref{bc6},\ref{a26} and \ref{a16} 
show the performance of the fit finite $N$ numbers
and their infinite $N$ extrapolated values. The $N=\infty$ estimate
differs from the data at the largest $N$ by $6\%$, $39\%$ and $10\%$
for $b_c$, $a_2$ and $a_1$ respectively. This amount of extrapolation
is roughly the same as that we had in the analysis of the synthetic
two dimensional data except for $a_2$, where it is around $20\%$ for
the synthetic data. All in all, although the extrapolations are quite
substantial, they are in line with expectations, and the two
dimensional study provides some confidence in the validity of the
infinite $N$ numbers we obtained in three dimensions.  The sample case
we show is typical of other loops we have analyzed. Always, the
determination of $b_c(L,\infty)$ is the most reliable. Next in terms
of reliability is the determination of $a_1(L,\infty)$. The
determination of $a_2(L,\infty)$ is the least reliable, perhaps
because of an amplification of small errors in the determination of
$b_c(L,\infty)$.

As a cross-check of the determination of the infinite $N$ numbers we 
analyze the $6^2$ loops also using the nonlinear simultaneous three parameter 
fit we described earlier.  We compare our three target numbers, $b_c(L,\infty)$ 
$a_1(L,\infty)$  and $a_2 (L,\infty)$ obtained 
in the nonlinear simultaneous fit to those obtained in the 
linear method based on  $\Omega$. 

It is only the infinite $N$ values that have to agree within errors, since
finite $N$ effects will differ in the two methods. We do not make a great 
effort to estimate the errors in the nonlinear fit, as it is used
only for a general consistency test. We observe a dependence on the
ranges we use which produces systematic errors that are larger 
than the statistical ones. It is this dependence on ranges we eliminate in
large measure (not completely though, as we need a range of $b$-values
for interpolation purposes, as explained) in the linear fitting method, based on 
$\Omega$. But, one may worry that focusing on too narrow a range in
$b$ can do more harm than good. This is the intuitive reason for our 
carrying out this consistency test. It goes above the usual feeling that 
nonlinear multi-parameter fits are less reliable then sequential linear fits.  

The results from the nonlinear simultaneous fit are shown 
by solid triangles in Figures 
\ref{bc6},\ref{a26} and \ref{a16}. The open triangles show
the performance of the fit versus $N$ and also show the
extrapolated values at infinite $N$.
As expected, they do not agree at finite $N$, 
but there is reasonable agreement on the extrapolated values at $N=\infty$. 
This assures us that focusing on the
region at $y=0$ in the linear method based on $\Omega$ 
does not present any dangers to reliability.  

The method used for fitting the parameters, as we have explained before, deals
with one parameter at a time and indirectly 
uses linear fits. This is our main method and it gave three numbers 
for each $V,L,N$, which are summarized in Table~\ref{tab1}

\subsection{Finite volume effects.}

As explained, one needs to test for contamination from finite volume effects,
even though there is large $N$ reduction promising an eventual lack of 
sensitivity to finite volume effects. It is not that the true infinite $N$ 
values are suspected of being dependent on torus size: The point is that the 
estimates we get for the infinite $N$ values are obtained by extrapolation from 
a set of finite $N$ values. Each one of these finite $N$ numbers does have a 
torus size dependence. Fitting to this finite set of values at finite $N$'s will 
produce best fit parameters that will depend on torus size too. The parameter 
giving the coefficient of $\frac{1}{N^0}$ will have some torus size dependence 
too, which would get weaker as more data at higher values of $N$ is made 
available. The coefficients of sub-leading terms of the form $\frac{1}{N^k}, k>0$ 
will have a dependence on torus size that is not supposed to disappear when data 
at higher values of $N$ is made available. In any finite set of data, the 
coefficient of $\frac{1}{N^0}$ will compensate, by some torus size dependence, 
for the absence of data at even higher values of $N$. 

We have tested for contamination from finite volume effects 
in two cases: We compare in Figures \ref{bcv},\ref{a2v}
and \ref{a1v} the results obtained using
a $4^3$ and a $8^3$ volume for a $3^2$ loop
and $3^3$, $4^3$ and $6^3$ volumes for a $2^2$ loop.  
We see that only smaller values of $N$ are affected. 
Also,  
the effect is stronger on the $3^2$ loop, which makes sense 
since $3^2$ loop on a $4^3$ volume 
ought  to be more contaminated then a $2^2$ loop on a $3^3$ volume. 
However, the main conclusion is positive:  our infinite $N$ values
are safe at our level of accuracy from finite volume contamination.  

\subsection{Extrapolation to infinite $L$ -- continuum extrapolation.}

The transition is a continuum feature; therefore, all the values of $L$
represent the same critical loop of a physical size $l_c$. In three dimensions
$b$ has dimensions of length, therefore $\frac{L}{b_c(L)}$ needs to approach
a finite limit. This limit is to be approached with corrections dictated by 
renormalization theory. Although the action generates only dimension two 
corrections, the observable has also dimension one corrections, and therefore all 
our fits are to three terms in a series in $\frac{1}{L}$ for $\frac{b_c}{L}$,
$a_1$ and $a_2$. 
Higher $L$ values run the danger of finite
volume contamination and therefore we avoided producing them. There also is a 
cost factor involved since
the higher $L$ is the larger the lattice $b$ is and 
consequently the larger the lattice volume has to be
in order to stay in the right phase, 0c. As usual, computer time will eventually
grow linearly with $L^3$.  

One way to check for consistency is to redo the fits by replacing the values 
$b_c(L)$ for each $L$ by their mean field, or tadpole, improved values:
\be
b^I (b) = b e(b)
\ee
where $e(b)$ is the average of $\frac{{\rm Tr} (U_p)}{N}$ for 
unsmeared parallel transporters round plaquettes $p$. It is known  that
a large fraction of $\frac{1}{b^2}$ corrections get absorbed when using
$b^I$ instead of $b$ as an extrapolation parameter to continuum. If our
continuum extrapolation method is to be more than 
merely asserted, it should give almost the same numbers in
the continuum limit when using either extrapolation method. However,
when using mean field improvement we should see 
less fractional variability as function of $L$. These features were indeed 
observed; thus our continuum extrapolation has passed a self-consistency check, 
which admittedly is somewhat heuristic.    

Figures \ref{bca},\ref{a2a} and \ref{a1a} show how the continuum limit
is approached.  The $2\times 2$ loop is probably on too coarse a
lattice and we perform fits which include and fits which exclude this
loop. Figure \ref{bca} shows the results for $\frac{b_c(L,\infty)}{L}$
and $\frac{{b_c}_I(L,\infty)}{L}$. The extrapolated value for both
data sets are consistent with each other but there is a significant
difference in the extrapolated value with and without the $2\times 2$
loop.  The $\chi^2$ value indicates that probably the lattice spacing
for the $2^2$ loops becomes too large. 
We therefore quote $0.113(6)$ as the continuum value for $b_c/l$
at $f=0.03$. The extrapolated result for $a_2$ has large errors as
expected and all we can quote is $5.8(2.3)$. The result for $a_1$ is
consistent with $a_1$ being unity.

\section{Summary and Discussion.}

We hypothesized that the strong to weak coupling phase transition
in large $N$ QCD is in the same universality class in two, three
and four dimensions.
Our primary finding is a picture that is consistent with our hypothesis
in three dimensional Euclidean YM. Moreover, it seems that the parameter
$a_1$ is consistent with the value $1$, indicating that indeed the
phase transition is as simple as it only could be.  

One should keep in mind that numerical tests are never foolproof.  
Even if it turns out that some details do not work as the hypothesis 
we formulated predicts, there could be weaker forms of the
hypothesis that do hold. We believe that it is better to have some clear
hypothesis one is testing, than just trying to accumulate a large body
of numerical information and look for systematics later. We hope that 
our hypothesis, or a competing one, will be independently 
checked in perhaps other ways.  

\section{Plans for the future.}

The first problem for the future is to extend this
work to four dimensions. If everything works like in three dimensions
one can proceed to address the question of shape dependence. 

It would be useful to derive double scaling limits for observables other 
than
the characteristic polynomial and use those for carrying out numerical tests. 
In particular, the double scaling limit of the distribution of extremal
eigenvalues would hold the promise of providing easier and more
stringent numerical checks.

A related question has to do with matching the universal data in the transition 
region to the perturbative side. For this, smearing and the precise
form of the observable become important practical details.  
Intuitively, the Grassmann/fermion representation for the 
characteristic polynomial provides a framework that 
has better potential to be amenable to
standard renormalized perturbation theory than whatever 
framework would be able to handle extremal eigenvalues.  
The  reason is that the Grassmann/fermion 
representation provides local expressions in spacetime, but it is hard to
imagine a space-time local approach for handling a double scaling limit
for the distribution of extremal eigenvalues.  
Nevertheless, exact results about extremal eigenvalues would be useful, only, 
one may need to match (this time within the matrix model) 
the parametric behavior of extremal eigenvalues
(assuming one can be obtained) to the 
description of the average characteristic polynomial 
we have been working with in this paper. 

The concept of an extremal eigenvalue is less natural for a unitary matrix
than it is for a hermitian one. In the unitary case, one 
would define ``extremal'' by the eigenvalue on the unit circle that is closest
to -1, the distance being measured by the shortest arc connecting the 
eigenvalue to -1 round the unit circle. Perhaps a more natural alternative is to
consider the probability distribution of the largest gap, where the
``gaps'' are measured by arcs connecting two consecutive eigenvalues round the 
circle.

\acknowledgments

R.N. acknowledge partial support by the NSF under grant number
PHY-055375.  H. N. acknowledges partial support by the DOE under grant
number DE-FG02-01ER41165 at Rutgers University and a Humboldt
Foundation prize.  The stay associated with the prize, at Humboldt
University Berlin, was very pleasant and H. N. is grateful to Ulli
Wolff and his computational physics group for their hospitality during
two long-term stays in Berlin.  Comments made by J. Feinberg, during
an extended visit at Rutgers, are also gratefully acknowledged.

\clearpage

{\footnotesize
\TABLE{
\begin{tabular}{cccccccc}
$V$ & $L$ & $N$ & $b_c(L,N)/L$  & ${b_c}_I(L,N)/L$& $a_2(L,N)$ 
& ${a_2}_I(L,N)$ & $a_1(L,N)$ \cr
\hline
$3^3$ & 2 & 17 & 0.2564(3) & 0.1578(4) & 1.69(3) & 0.91(2) & 0.831(1) \cr
$3^3$ & 2 & 23 & 0.2506(3) & 0.1508(3) & 1.88(4) & 0.97(2) & 0.846(1) \cr
$3^3$ & 2 & 29 & 0.2483(3) & 0.1479(3) & 2.01(6) & 1.03(3) & 0.855(1) \cr
$3^3$ & 2 & 37 & 0.2463(3) & 0.1454(3) & 2.19(5) & 1.09(2) & 0.863(2) \cr
$3^3$ & 2 & 41 & 0.2457(3) & 0.1447(3) & 2.12(5) & 1.06(3) & 0.866(2) \cr
$3^3$ & 2 & 47 & 0.2444(3) & 0.1433(3) & 2.18(6) & 1.09(3) & 0.872(2) \cr
$4^3$ & 2 & 17 & 0.2568(2) & 0.1581(2) & 1.61(2) & 0.87(1) & 0.832(0) \cr
$4^3$ & 2 & 23 & 0.2515(2) & 0.1517(2) & 1.82(2) & 0.95(1) & 0.845(1) \cr
$4^3$ & 2 & 29 & 0.2483(2) & 0.1479(2) & 1.97(3) & 1.00(1) & 0.856(1) \cr
$4^3$ & 2 & 37 & 0.2461(2) & 0.1452(3) & 2.08(4) & 1.04(2) & 0.864(1) \cr
$4^3$ & 2 & 41 & 0.2452(2) & 0.1441(3) & 2.12(4) & 1.05(2) & 0.868(1) \cr
$4^3$ & 2 & 47 & 0.2447(2) & 0.1435(2) & 2.28(5) & 1.13(2) & 0.871(1) \cr
$6^3$ & 2 & 17 & 0.2567(1) & 0.1579(1) & 1.60(1) & 0.87(1) & 0.832(0) \cr
$6^3$ & 2 & 23 & 0.2514(1) & 0.1516(1) & 1.79(1) & 0.94(1) & 0.846(0) \cr
$6^3$ & 2 & 29 & 0.2485(1) & 0.1481(1) & 1.94(2) & 0.99(1) & 0.855(0) \cr
$6^3$ & 2 & 37 & 0.2460(1) & 0.1451(1) & 2.07(2) & 1.03(1) & 0.865(1) \cr
$6^3$ & 2 & 41 & 0.2454(1) & 0.1443(1) & 2.14(2) & 1.06(1) & 0.868(1) \cr
$6^3$ & 2 & 47 & 0.2445(1) & 0.1433(1) & 2.24(2) & 1.11(1) & 0.872(1) \cr
$4^3$ & 3 & 17 & 0.2309(3) & 0.1695(3) & 1.39(2) & 0.98(1) & 0.838(0) \cr
$4^3$ & 3 & 23 & 0.2258(3) & 0.1641(3) & 1.45(2) & 1.00(2) & 0.854(1) \cr
$4^3$ & 3 & 29 & 0.2228(2) & 0.1608(2) & 1.45(2) & 1.00(1) & 0.866(1) \cr
$4^3$ & 3 & 37 & 0.2197(2) & 0.1575(2) & 1.50(3) & 1.03(2) & 0.878(1) \cr
$4^3$ & 3 & 41 & 0.2185(2) & 0.1564(2) & 1.55(2) & 1.06(1) & 0.883(1) \cr
$4^3$ & 3 & 47 & 0.2177(2) & 0.1554(2) & 1.59(4) & 1.08(2) & 0.888(1) \cr
$8^3$ & 3 & 17 & 0.2344(1) & 0.1731(1) & 1.27(1) & 0.90(0) & 0.838(0) \cr
$8^3$ & 3 & 23 & 0.2268(1) & 0.1652(1) & 1.36(1) & 0.95(0) & 0.855(0) \cr
$8^3$ & 3 & 29 & 0.2228(1) & 0.1609(1) & 1.44(1) & 0.99(1) & 0.867(0) \cr
$8^3$ & 3 & 37 & 0.2199(1) & 0.1578(1) & 1.51(1) & 1.03(1) & 0.878(0) \cr
$8^3$ & 3 & 41 & 0.2188(1) & 0.1566(1) & 1.54(1) & 1.05(1) & 0.883(0) \cr
$8^3$ & 3 & 47 & 0.2176(1) & 0.1554(1) & 1.54(4) & 1.05(2) & 0.889(0) \cr
$8^3$ & 4 & 17 & 0.2189(1) & 0.1741(1) & 1.19(1) & 0.93(1) & 0.839(0) \cr
$8^3$ & 4 & 23 & 0.2115(1) & 0.1665(1) & 1.30(1) & 1.00(1) & 0.856(0) \cr
$8^3$ & 4 & 29 & 0.2075(1) & 0.1624(1) & 1.34(1) & 1.02(1) & 0.868(0) \cr
$8^3$ & 4 & 37 & 0.2043(1) & 0.1591(1) & 1.41(1) & 1.07(1) & 0.880(0) \cr
$8^3$ & 4 & 41 & 0.2031(1) & 0.1578(1) & 1.42(1) & 1.08(1) & 0.885(0) \cr
$8^3$ & 4 & 47 & 0.2021(1) & 0.1568(1) & 1.46(1) & 1.10(1) & 0.890(0) \cr
$8^3$ & 5 & 17 & 0.2054(1) & 0.1701(1) & 1.17(1) & 0.96(1) & 0.839(0) \cr
$8^3$ & 5 & 23 & 0.1986(1) & 0.1631(1) & 1.26(1) & 1.02(1) & 0.856(0) \cr
$8^3$ & 5 & 29 & 0.1949(3) & 0.1594(3) & 1.32(2) & 1.06(2) & 0.869(1) \cr
$8^3$ & 5 & 37 & 0.1919(1) & 0.1563(1) & 1.37(1) & 1.10(1) & 0.880(0) \cr
$8^3$ & 5 & 41 & 0.1907(1) & 0.1551(1) & 1.40(4) & 1.12(3) & 0.887(2) \cr
$8^3$ & 5 & 47 & 0.1896(1) & 0.1540(1) & 1.45(2) & 1.16(1) & 0.892(1) \cr
$8^3$ & 6 & 17 & 0.1944(2) & 0.1652(2) & 1.24(1) & 1.04(1) & 0.839(0) \cr
$8^3$ & 6 & 23 & 0.1889(2) & 0.1597(2) & 1.26(1) & 1.06(1) & 0.856(0) \cr
$8^3$ & 6 & 29 & 0.1854(1) & 0.1561(1) & 1.28(1) & 1.07(1) & 0.868(0) \cr
$8^3$ & 6 & 37 & 0.1826(1) & 0.1532(1) & 1.34(1) & 1.11(1) & 0.881(0) \cr
$8^3$ & 6 & 41 & 0.1818(2) & 0.1524(2) & 1.39(3) & 1.15(2) & 0.885(1) \cr
$8^3$ & 6 & 47 & 0.1805(1) & 0.1512(1) & 1.42(2) & 1.18(2) & 0.892(1) \cr
\hline
\end{tabular}
\caption{\label{tab1} The results for the
parameters matching lattice data to the double scaling function at different 
values of $V$, $L$ and $N$.}}
}

\clearpage
\FIGURE{
\epsfig{file=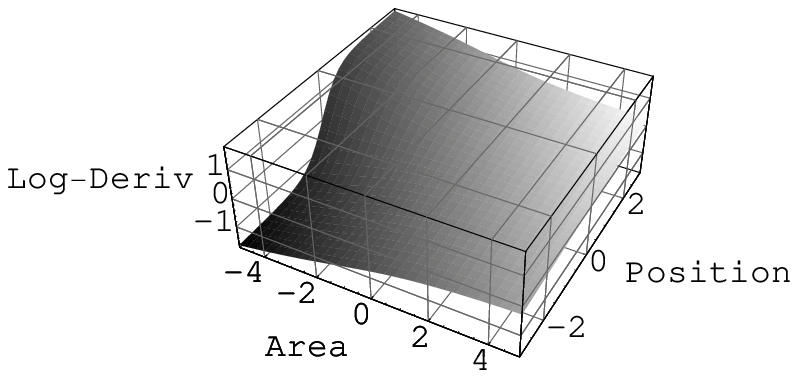, height=2.8in}
\caption{ Plot of the logarithmic derivative of $\zeta$ with
respect to $\xi$ for different values of the ``area'' ($\alpha$)
and ``position'' ($\xi$). One sees the smooth remnant
of the singularity at $\xi=0$ and its dependence on the area. }
\label{smoothfig}}

\FIGURE{
\epsfig{file=omegas.eps, height=2.8in }
\caption{Plot of $\Omega(\alpha)$ as a function of $\alpha$ 
showing the behavior of the scaled function. 
}
\label{omegas}}

\FIGURE{
\epsfig{file=omega.eps, height=2.8in }
\caption{Plot of $\Omega(b,N)$ as a function of $b$ for several $N$. 
$N=17,23,29,37,41,47$ are the values in the plot and they
gradually approach the critical behavior at $N=\infty$.
The dashed line is the value of $\Omega$ at the critical point
obtained using the double scaling limit.
}
\label{omegan}}

\FIGURE{
\epsfig{file=bc.eps, height=2.8in }
\caption{Plot of $b_c(N)$ as a function of $\frac{1}{N}$. 
}
\label{bcn}}

\FIGURE{
\epsfig{file=a2.eps, height=2.8in }
\caption{Plot of $a_2(N)$ as a function of $\frac{1}{\sqrt{N}}$. 
}
\label{a2n}}

\FIGURE{
\epsfig{file=a1.eps, height=2.8in }
\caption{Plot of $a_1(N)$ as a function of $\frac{1}{\sqrt{N}}$. 
}
\label{a1n}}

\FIGURE{
\epsfig{file=energy.eps, height=2.8in }
\caption{Plot of the average plaquette, $e(b)$,
 as a function of $b$. 
}
\label{plaq}}

\FIGURE{
\epsfig{file=fcritvsl.eps, height=2.8in }
\caption{Plot of the critical value of the smearing parameter
 as a function of the size of the loop. 
}
\label{fcrit}}

\FIGURE{
\epsfig{file=omega47.eps, height=2.8in }
\caption{Plot of the $\Omega(b,N,L)$ as a function
of $b$ for $N=47$ and $L=3$
on a $4^3$ lattice.  
}
\label{omega47}}

\FIGURE{\epsfig{file=bc6.eps, height=2.8in }
\caption{Plot of the $\frac{b_c(L,N)}{L}$ as a function
of $1/N$ for $L=6$ 
on a $8^3$ lattice.  
}
\label{bc6}}

\FIGURE{
\epsfig{file=a26.eps, height=2.8in }
\caption{Plot of the $a_2(L,N)$ as a function
of $1/\sqrt{N}$ for $L=6$ 
on a $8^3$ lattice.  
}
\label{a26}}

\FIGURE{
\epsfig{file=a16.eps, height=2.8in }
\caption{Plot of the $a_1(L,N)$ as a function
of $1/\sqrt{N}$ for $L=6$ 
on a $8^3$ lattice.  
}
\label{a16}}

\FIGURE{
\epsfig{file=bcv.eps, height=2.8in }
\caption{Plot of the $\frac{b_c(L,N)}{L}$ as a function
of $N$ for different lattices sizes and $L=2,3$. 
}
\label{bcv}}

\FIGURE{
\epsfig{file=a2v.eps, height=2.8in }
\caption{Plot of the $a_2(L,N)$ as a function
of $N$ for different lattices sizes and $L=2,3$.
}
\label{a2v}}

\FIGURE{
\epsfig{file=a1v.eps, height=2.8in }
\caption{Plot of the $a_1(L,N)$ as a function
of $N$ for different lattices sizes and $L=2,3$. 
}
\label{a1v}}

\FIGURE{
\epsfig{file=bca.eps, height=2.8in }
\caption{Plot of the $\frac{b_c(L,\infty)}{L}$ as a function
of $1/L$. 
}
\label{bca}}

\FIGURE{
\epsfig{file=a2a.eps, height=2.8in }
\caption{Plot of $a_2(L,\infty)$ as a function
of $1/L$.}\label{a2a}}

\FIGURE{
\epsfig{file=a1a.eps, height=2.8in }
\caption{Plot of $a_1(L,\infty)$ as a function
of $1/L$.}\label{a1a}}

\end{document}